\documentclass[amsmath,amssymb,twocolumn,prc,showpacs,nofootinbib,
floatfix,superscriptaddress]{revtex4}
\usepackage{graphics}
\usepackage{dcolumn}
\usepackage{bm}
\newcommand{\penelope}[0]{{\scshape Penelope}}

\newcommand{\apjs}{Astrophys. J. Suppl. Ser.}
\newcommand{\aap}{Astron. Astrophys.}
\newcommand{\apjl}{Astrophys. J. Lett.}
\def\prc{ Phys.\ Rev.\ C }
\begin{document}
\title{Absolute determination of the $^{22}$Na($p,\gamma$)$^{23}$Mg reaction rate in novae} 
\author{A. L. Sallaska} 
\altaffiliation{Present address:  Department of Physics and Astronomy, University of North Carolina at Chapel Hill, Chapel Hill, NC 27599-3255, USA}
\affiliation{Department of Physics, University of Washington,
Seattle, WA 98195-1560, USA}
\email[Corresponding author:  ]{sallaska@physics.unc.edu}
\author{C. Wrede}
\affiliation{Department of Physics, University of Washington,
Seattle, WA 98195-1560, USA}
\author{A. Garc\'{\i}a}
\affiliation{Department of Physics, University of Washington,
Seattle, WA 98195-1560, USA}
\author{D. W. Storm}
\affiliation{Department of Physics, University of Washington,
Seattle, WA 98195-1560, USA}
\author{T. A. D. Brown}
\altaffiliation{Present address:  Mary Bird Perkins Cancer Center, Baton Rouge, LA 70809}
\affiliation{Department of Physics, University of Washington,
Seattle, WA 98195-1560, USA}
\author{C. Ruiz}
\affiliation{TRIUMF, Vancouver, BC V6T 2A3, Canada}
\author{K. A. Snover}
\affiliation{Department of Physics, University of Washington,
Seattle, WA 98195-1560, USA}
\author{D. F. Ottewell}
\affiliation{TRIUMF, Vancouver, BC V6T 2A3, Canada}
\author{L. Buchmann}
\affiliation{TRIUMF, Vancouver, BC V6T 2A3, Canada}
\author{C. Vockenhuber}
\altaffiliation{Present address:  ETH Laboratory of Ion Beam Physics, Zurich, Switzerland CH-8093}
\affiliation{TRIUMF, Vancouver, BC V6T 2A3, Canada}
\author{D. A. Hutcheon}
\affiliation{TRIUMF, Vancouver, BC V6T 2A3, Canada}
\author{J. A. Caggiano}
\altaffiliation{Present address:  LLNL, Livermore, CA 94550}
\affiliation{TRIUMF, Vancouver, BC V6T 2A3, Canada}
\author{J. Jos\'{e}}
\affiliation{Departament F\'{i}sica i Enginyeria Nuclear (UPC) and Institut d'Estudis Espacials de Catalunya (IEEC) , E-08034 Barcelona, Spain}
\date{\today}
\begin{abstract}
Gamma-ray telescopes in orbit around the Earth are searching for evidence of the elusive radionuclide $^{22}$Na produced in novae.  Previously published uncertainties in the dominant destructive reaction, $^{22}$Na($p,\gamma)^{23}$Mg, indicated new measurements in the proton energy range of 150 to 300 keV were needed to constrain predictions.  We have measured the resonance strengths, energies, and branches directly and absolutely by using protons from the University of Washington accelerator with a specially designed beamline, which included beam rastering and cold vacuum protection of the $^{22}$Na implanted targets.  The targets, fabricated at TRIUMF-ISAC, displayed minimal degradation over a $\sim$ 20 C bombardment as a result of protective layers.  We avoided the need to know the absolute stopping power, and hence the target composition, by extracting resonance strengths from excitation functions integrated over proton energy.  Our measurements revealed that resonance strengths for $E_p$  = 213, 288, 454, and 610 keV are stronger by factors of 2.4 to 3.2 than previously reported.  Upper limits have been placed on proposed resonances at 198-, 209-, and 232-keV.  These substantially reduce the uncertainty in the reaction rate.  We have re-evaluated the $^{22}$Na($p,\gamma)$ reaction rate, and our measurements indicate the resonance at 213 keV makes the most significant contribution to $^{22}$Na destruction in novae.  Hydrodynamic simulations including our rate indicate that the expected abundance of $^{22}$Na ejecta from a classical nova is reduced by factors between 1.5 and 2, depending on the mass of the white-dwarf star hosting the nova explosion.
\end{abstract}
\pacs{29.30.Kv, 29.38.Gj, 26.30.Ca, 27.30.+t}
\maketitle
\section{Motivation}

A classical nova is the consequence of thermonuclear runaway on the surface of a white-dwarf star that is accreting hydrogen-rich material from its partner in a binary system.  Such novae are ideal sites for the study of explosive nucleosynthesis because the observational~\cite{starrfield}, theoretical~\cite{starrfield97,jose98}, and nuclear-experimental~\cite{iliadis02,josenew} aspects of their study are each fairly advanced. In particular, due to the relatively low peak temperatures in nova outbursts ($0.1 < T < 0.4$ GK), most of the nuclear reactions involved are not too far from the valley of beta stability to be studied in the laboratory, and the corresponding thermonuclear reaction rates are mostly based on experimental information~\cite{iliadis02}.

An example of a gamma-ray emitter produced in novae is $^{26}$Al (t$_{1/2}$ $\sim$ $7.7\times 10^5$ years, $E_\gamma$ = 1.809 MeV).  Indeed, this isotope has been observed in the Galaxy~\cite{diehl}, but its long half life precludes the identification of its progenitor, and novae are only expected to make a secondary contribution to its Galactic abundance~\cite{jose, ru06prl}.  Other gamma-ray emitters can provide more direct constraints on nova models~\cite{clayton}.  An example is $^{22}$Na (t$_{1/2}$ = 2.603 years,  $E_\gamma$ = 1.275 MeV), which has not yet been observed in the Galaxy.  Unlike $^{26}$Al, the relatively short half life of $^{22}$Na restricts it to be localized near its production site.  Novae also could be the principal Galactic sites for the production of $^{22}$Na, making $^{22}$Na an excellent nova tracer.  An observational upper limit of $3.7 \times 10^{-8}$ M$_{\odot}$ was set on the $^{22}$Na mass in ONe nova ejecta with the COMPTEL telescope onboard the CGRO~\cite{iyudin}.  Currently, the maximum $^{22}$Na mass ejected using ONe nova models is an order of magnitude below this limit~\cite{jose,bishop} and corresponds to a maximum detection distance of 500 parsecs using an observation time of $10^6$ s with the spectrometer SPI onboard the currently-deployed INTEGRAL mission~\cite{hernanz}.  This suggests we are now on the verge of being able to detect this signal.  In addition, Ne isotopic ratios in some meteoritic presolar graphite grains imply the \emph{in-situ} decay of $^{22}$Na produced by nucleosynthesis in novae~\cite{nittler,jose04} and/or supernovae~\cite{amari}.  

It is important to reduce uncertainties in the rates of key reactions that are expected to affect the production of $^{22}$Na so that accurate comparisons can be made between observations and models~\cite{jose}.  For example, the production of $^{22}$Na in novae depends strongly on the thermonuclear rate of the $^{22}$Na($p,\gamma)^{23}$Mg reaction~\cite{hix,jose,iliadis02}, which consumes $^{22}$Na.  The thermonuclear $^{22}$Na($p,\gamma$) reaction rate in novae is dominated by narrow, isolated resonances with laboratory proton energies $E_p \lesssim 300$ keV.  Consequently, the rate is dependent on the energies and strengths of these resonances, which have been investigated both indirectly and directly in the past.  Indirect information on potential $^{22}$Na($p,\gamma$) resonances has been derived from measurements of the $^{24}$Mg($p,d$)~\cite{kubono}, $^{25}$Mg($p,t$)~\cite{nann}, and $^{22}$Na($^3$He,$d$)~\cite{schmidt} reactions, and from the beta-delayed proton- and gamma-decays of $^{23}$Al~\cite{tighe,per,iacob}.  The first published attempt to measure the $^{22}$Na($p,\gamma$) reaction directly employed a chemically prepared, radioactive $^{22}$Na target and produced only upper limits on the resonance strengths~\cite{gorres}.  A measurement contemporary to Ref.~\cite{gorres} in the range $E_p > 290$ keV by Seuthe \emph{et al.} employed ion-implanted $^{22}$Na targets~\cite{seuthe}, resulting in the first direct observation of resonances and the only absolute measurement of resonance strengths.  Later, Stegm\"{u}ller \emph{et al.}~\cite{stegmuller} discovered a new resonance at 213 keV and determined its strength relative to the strengths from Ref.~\cite{seuthe}.  More recently, a new level in $^{23}$Mg ($E_x$ = 7770 keV) has been discovered using the $^{12}$C($^{12}$C,$n\gamma$)~\cite{jenkins} reaction.  This level corresponds to a $^{22}$Na($p,\gamma$) laboratory proton energy of 198 keV, and the authors of Ref.~\cite{jenkins} proposed that this potential resonance could dominate the $^{22}$Na($p,\gamma$) reaction rate at nova temperatures.

We have measured the energies, strengths, and branches of known resonances~\cite{seuthe,stegmuller} and searched for proposed resonances~\cite{tighe,per,iacob,jenkins} in the energy range $E_p$  $\sim$ 195 to 630 keV.  The measurements were performed at the Center for Experimental Nuclear Physics and Astrophysics of the University of Washington with ion-implanted $^{22}$Na targets prepared at TRIUMF-ISAC.  Thanks to evaporated protective layers~\cite{tom}, the targets exhibited little to no degradation over $\sim$ 20 C of bombardment.  Using mainly the strengths and energies obtained in this work together with supplemental information from other work~\cite{seuthe,iliadispreprint3}, we have re-evaluated the thermonuclear reaction rate of $^{22}$Na($p,\gamma$), and full hydrodynamic simulations have been performed to estimate the effect of the new rate on the flux of $^{22}$Na from novae.  This article is a detailed presentation of our experiment, its results, and their implications, complementing our previous reports~\cite{sallaska,thesis}.

\section{Experimental Setup}

We measured $^{22}$Na($p,\gamma$) resonances directly by bombarding implanted $^{22}$Na targets with protons from a tandem Van de Graaff accelerator.  High currents ($\sim$ 45 $\mu$A) at lab energies ranging from 150 to 700 keV were achieved with a terminal ion source.  

\subsection{Strategy}

The number of reactions, $N_{\rm prod}$, produced by a beam of incident particles with areal density 
$dN_b/dA$ on a target with areal density $dN_T/dA$ is given by

\begin{equation}
N_{\rm prod} = \sigma \int \frac{dN_T}{dA}\frac{dN_b}{dA}\, dA,
\label{first}
\end{equation}
where $\sigma$ is the cross section.  Conventional methods employ a small-diameter beam that impinges on a large-area target, where the target density is nearly uniform.  However, this technique can lead to target damage in cases where large beam currents are used, and there is a long history of differing results on resonance strengths that have been attributed to target instabilities~\cite{long1,long2,long3,long4}.  We designed our experiment closer to the opposite limit, similar to Ref.~\cite{kurt}, where the beam was swept over an area larger than the full extent of the target with a rastering device.  In the limit of uniform beam density over the target area, Eq.~\ref{first} becomes

\begin{equation}
N_{\rm prod} = \sigma N_T \frac{dN_b}{dA}.\label{second}
\end{equation}
This method requires knowledge of only the {\em total} number of target atoms and, thus, is not very sensitive to target non-uniformities.  On the other hand, this method also requires a determination of the beam density.  The yield is given by
\begin{equation}
Y = \sigma N_T \rho_b,\label{third}
\end{equation}
where $\rho_b =  \frac{dN_b}{dA}/(Q/e) $ is a beam density normalized to the accumulated charge, $Q$.

In addition, we determined the {\em integrated yield} of the excitation function over the beam energy, minimizing uncertainties associated with the energy loss in the target and beam energy distribution, which can be substantial in determinations using only the yield at a particular energy.  The latter method, which was used in Ref.~\cite{seuthe}, depends on knowing the energy loss in the target, the target stochiometry and uniformity, and often assumes stable target conditions, which are unlikely in experiments with currents of tens of microamps, such as ours.

Beginning with Eq.~\ref{third} (see, for example, Ch. 4 of Ref.~\cite{iliadisbook}), the integrated yield for a finite-thickness target is given by
\begin{equation}
\int Y \, dE = 2 \pi^2 \lambdabar^2 \frac{m+M}{M} \,N_T \, \rho_b \, \omega\gamma, \label{main}
\end{equation}
where $\int Y \, dE$ is the integral over the laboratory beam energy $E$ with a range spanning the resonance, $\lambdabar$ is the reduced de Broglie wavelength in the center of mass, $m$ and $M$ are the projectile and target mass, respectively, and $\omega\gamma$ is the resonance strength. 
\subsection{Targets}\label{target}

Our $^{22}$Na targets were produced by ion implantation, which yields isotopically pure targets and avoids complications with chemical fabrication.  Each target was made at TRIUMF-ISAC by implanting a 10-nA, 30-keV $^{22}$Na$^+$ ion beam into a rectangular OFHC copper substrate with dimensions 3 mm $\times$ 19 mm $\times$ 25 mm.  The beam was rastered over a 5-mm collimator such that at the raster extreme only 5\% of the beam remained on target, thereby creating a nearly uniform density.  The setup included electron suppression with $-300$ V and a liquid nitrogen-cooled cold trap with a vacuum pressure in the range of $6\times 10^{-8}$ to $2\times 10^{-7}$ torr.  Charge integration was monitored throughout the implantation process, which took roughly 24 hours per target with a peak $^{22}$Na current of $\sim 15$ nA.  

Initially, two test targets, \#1 and \#2, were implanted with activities of $\sim$ 300 and 185 $\mu$Ci, respectively.  As target degradation can be quite problematic, we carried out a program~\cite{tom} to determine the ideal combination of implantation energy, substrate, and possible protective layer by bombarding $^{23}$Na targets implanted under similar conditions.  Using the conclusions of Ref.~\cite{tom}, two additional $\sim$ 300 $\mu$Ci $^{22}$Na targets, \#3 and \#4, were implanted with the same parameters but included a 20-nm protective layer of chromium, deposited by vacuum evaporation after implantation.  A small rise in temperature of the target was observed during the evaporation; however, a survey of the apparatus showed no residual activity from diffusion of $^{22}$Na out of the target.  All $^{22}$Na data presented were taken on the chromium-covered targets, with the exception of the 232-keV resonance measurement that employed target \#1.

To explore the transverse location of the implanted $^{22}$Na, the beta activity was scanned with a Geiger counter behind a 6-mm thick brass shield.   A 3-mm diameter hole in the center allowed transmission of the beta particles.  This measurement confirmed that the $^{22}$Na was confined within a 5-mm diameter circle and determined the position of the activity relative to the center of the substrate.  Although this method was not very sensitive to radially-dependent non-uniformity, it did verify axial symmetry.  Thanks to this information and the extreme rastering of the $^{22}$Na beam, we believe the targets were quite uniform, although our method does not require this.  


\subsection{Chamber and Detectors}

\begin{figure}
\hfil\scalebox{.33}{\includegraphics{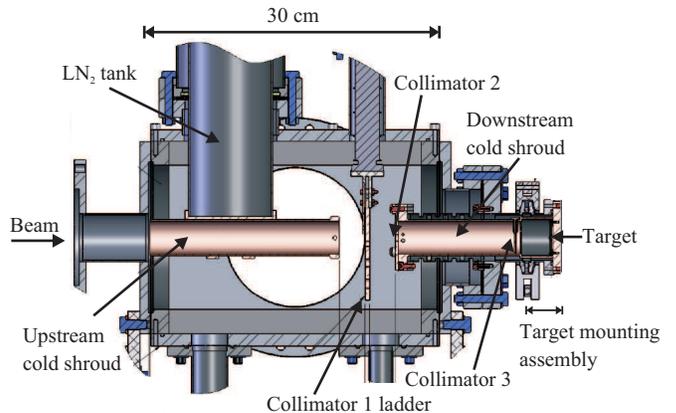}}\hfil
\caption{(color online)  Side view of chamber cross section.  Copper braids connecting the upstream and downstream cold shrouds have been omitted for clarity.}
\label{chamber}
\end{figure}

\begin{figure}
\hfil\scalebox{.5}{\includegraphics{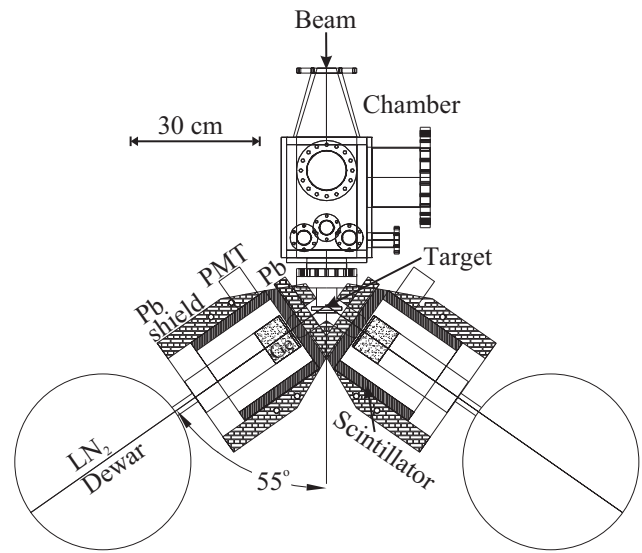}}\hfil
\caption{Top view of detector setup. The PMT shown is for the annular scintillator only; the planar scintillator PMT has been omitted for clarity. }
\label{setup}
\end{figure}%

Figs.~\ref{chamber} and~\ref{setup} illustrate our chamber and detector system, respectively.  The main features of the chamber included its dual liquid-nitrogen cooled cold shroud system, three collimators, and water-cooled target mount.  The cold shroud isolated the radioactive target to prevent the contamination of the upstream beamline with $^{22}$Na and also helped maintain a clean environment near the target, suppressing carbon buildup.  During data collection, the pressure in the chamber was in the range of (1-2)$\times 10^{-7}$ torr.  The end of the downstream cold shroud surrounded our target substrate; however, because it was the farthest away from the liquid-nitrogen cold trap, it only reached a temperature of 125 K, whereas the upstream shroud reached 88 K.

The chamber had three sets of collimators.  The first, collimator 1, was a water-cooled, sliding ladder between the cold shrouds with 4-, 7-, and 8-mm diameter collimators.  The 8-mm collimator was used during $^{22}$Na($p,\gamma$) data acquisition.  Also on this ladder were electron-suppressed 1-mm and 3-mm diameter collimators with downstream beam stops for tuning.  Collimator 2, an 8-mm collimator, was 33 mm downstream of the ladder and was attached to the end of the downstream cold shroud.  It was followed by a 10-mm diameter cleanup collimator, collimator 3, located 122 mm farther downstream.  Each collimator was electrically isolated from the chamber to permit current monitoring.  

The target substrate was bolted to a copper backing flange and was directly cooled with deionized-water via a thin pipe coupled to the flange.  To minimize handling time in proximity to the radioactive targets, this assembly was then attached to a stainless steel coupler with a ISO flange on the chamber side.  Directly upstream of this assembly was a 30-mm long electron suppressor biased between $-150$ and $-300$ V.  During data collection, the current on target was monitored, and the charge was integrated and recorded.

Two sets of detector systems were positioned at $\pm$ 55$^\circ$ to the beam axis.  A top view of the setup, including chamber, shielding, detectors, and Dewars, is shown in Fig.~\ref{setup}.  Each system consisted of a high-purity 100\% germanium (HPGe) crystal Canberra model GR10024 encased in cosmic-ray anticoincidence shielding.  The detector angle was chosen to minimize effects due to angular anisotropy, as the Legendre polynomial $P_2(\cos\theta)$ is zero at $\pm$ 55$^\circ$.  The resolutions for each detector at 1.275 MeV were 4.4 keV and 7.4 keV (FWHM) with high rates ($^{22}$Na target present), and 2.2 keV and 3.0 keV (FWHM) with low rates (using residual $^{22}$Na activity with $^{22}$Na target removed).

In addition, because of the target radioactivity, 26 mm of lead shielding was placed between the target and detector system to reduce the event rate in the detectors to a few tens of kHz.  According to simulations, described in detail in Sec.~\ref{penelope}, the lead reduced the counting rate for the 511-keV gamma ray by a factor of 70, whereas the photopeak from 1275-keV gamma rays was reduced by a factor of 5.  The suppression ranged from factors of $\sim$ 3.5 to 4.5 for gamma rays with energies above $\sim$ 5 MeV.  Above 4 MeV, cosmic rays caused a continuum background.  In order to remove these unwanted signals, 25-mm thick annular and planar plastic scintillators encased in lead, as shown in Fig.~\ref{setup}, were used in anticoincidence with the germanium detectors and are discussed in detail in the next section.  The reduction of the $^{22}$Na($p,\gamma$) signal by this veto was negligible.   
\subsection{Data Acquisition}


In addition to the signals from the high-purity Ge detectors, the electronics also processed PMT signals from the scintillators.  In order to reduce the rate seen by the detectors due to target radioactivity, two techniques were used:  1) lead shielding was installed as described above, and 2) a high threshold was set (just below the strong 1275-keV $^{22}$Na line, so that the target activity could be monitored \emph{in-situ}).  

\begin{figure}
\hfil\scalebox{.35}{\includegraphics{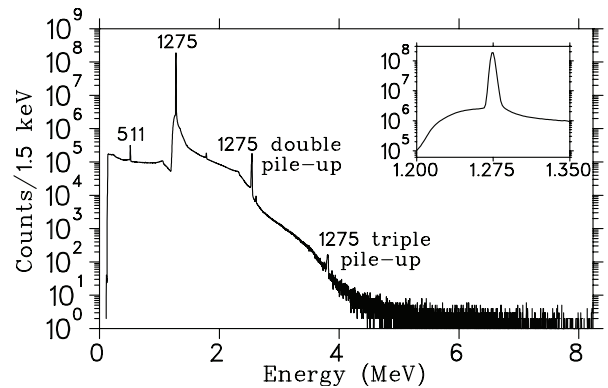}}\hfil
\caption{Sample background spectrum from one Ge detector, measured for $\sim$ 160 hours.  Inset shows the spectrum expanded around the 1275-keV gamma ray.}
\label{samplespectrum}
\end{figure}

The Ge-signal amplifiers (ORTEC 672) were operated in a pile-up-rejection mode (which typically rejected $\sim$ 40\% and introduced a dead time of 27 $\mu$s per pulse).  Signals from the two sets of detectors were converted into digital signals by ORTEC AD413A ADCs with fast FERAbus readout, which helped to reduce dead time.  A buffer module was used to minimize the communication with the computer via a CAMAC interface.  JAM, a JAVA-based data acquisition and analysis package for nuclear physics~\cite{jam}, was used to process the data.  All the NIM and CAMAC electronics modules were located in a temperature-controlled rack to minimize instabilities.  The raw rate in each detector was below 30 kHz, and the trigger rate was $\sim$ 4 kHz.  A sample background spectrum from one Ge detector is shown in Fig.~\ref{samplespectrum}.

Signals from the active anticoincidence shields and from the Ge detectors were used as a stop and start, respectively, in a Time-to-Amplitude-Converter (TAC).  If these two signals occurred within a set timing window, the resulting Ge signal was discarded.  



In order to determine the bounds of the TAC spectrum for data processing, the TAC signal corresponding to germanium detection energies between 4 and 6 MeV was extracted.  The timing gate was set on the prompt peak, which had a long tail to its left.  In this energy region, this anticoincidence system rejected $\sim$ 80\% of the cosmic-ray background signal.  The TAC signal for most high-energy cosmic rays from $\sim$ 6 to 11 MeV also falls within the set window.  The scintillator threshold was set well above 511 keV to avoid vetoes by annihilation radiation.  Self veto is possible with cascade gamma decays if one gamma ray is registered in the Ge and the other in the anticoincidence shield; this was examined by comparing the yields from raw singles spectra with those from the vetoed spectra, which agreed to better than 99\%.

\subsection{Detector Efficiency}

Detector photopeak  efficiencies were obtained by combining \penelope~\cite{penelope1,penelope2} simulations with direct measurements of gamma rays from $^{60}$Co and $^{24}$Na sources and from $^{27}{\rm Al}(p,\gamma)$ resonance measurements.  

The efficiency at 1332 keV was determined using a 31.51-nCi $^{60}$Co source.  Isotope Products~\cite{isotopeprod} produced the source and measured its decay intensity with an uncertainty of 1.7\% (99\% $C.L.$).  The geometry for the \penelope\ simulations was adjusted slightly to match the efficiency at this energy, as described in the following subsection.  The ratio of efficiencies from 1369 to 2754 keV was measured using a $^{24}$Na source ($t_{1/2}= 15$ hrs) fabricated at the University of Washington.  Since 1369 is very close to 1332 keV, we equated the efficiency at 1369 keV to the \penelope\ value, and then from this ratio we obtained the efficiency at 2754 keV.  

To extend the efficiency determination to higher energies, we measured the $^{27}$Al($p,\gamma$) reaction using a thick aluminum target and used the relative intensities of well-known gamma-ray branches from $E_p$ = 633 and 992~\cite{al1,al2}.  For the latter resonance, our coin target was also used.  We subtracted the below-resonance yield from the above-resonance yield to extract the net yield.  At $E_p = 992$ keV, the gamma rays of interest are at 1779, 4742, and 10762 keV.  Using the simulation to determine the efficiency at 1779 keV, our measurements, combined with the known branches, gave the efficiencies at the two higher energies.  At $E_p$ = 633 keV we measured the gamma rays at 7575 keV and 10451 keV.  As the simulation matched the value we had obtained at 10762 keV, we used the simulation for the 10451 keV value and the known branch to determine the efficiency at 7575 keV.  The agreement between measurement and simulation is discussed in the following subsection.

\subsubsection{\penelope\, Simulations}\label{penelope}


The geometry of our apparatus was modeled in the detailed Monte Carlo code \penelope.  The simulated germanium detector included the germanium crystal, cold finger, and carbon window, with all dimensions taken from the nominal specifications provided by Canberra~\cite{canberra}.   In addition, we included the 26-mm lead and 25-mm planar plastic scintillator in front of the detector.  Although the annular plastic scintillator was modeled, the annular lead was not, as it was not in the line of sight of the target.  The sodium source was a uniform 5-mm diameter disk centered on the copper substrate.  The copper backing mount was included, as was the aluminum plate supporting the water cooling system.  The water and its copper pipes inside the target mount were modeled, but the pipes that extended up and out from the mount were not, as they were thin and mostly out of the line of sight.  All components of the target mount were aligned with the beam, whereas the detector was at an angle of 55$^\circ$.  

\begin{figure}
\hfil\scalebox{.35}{\includegraphics{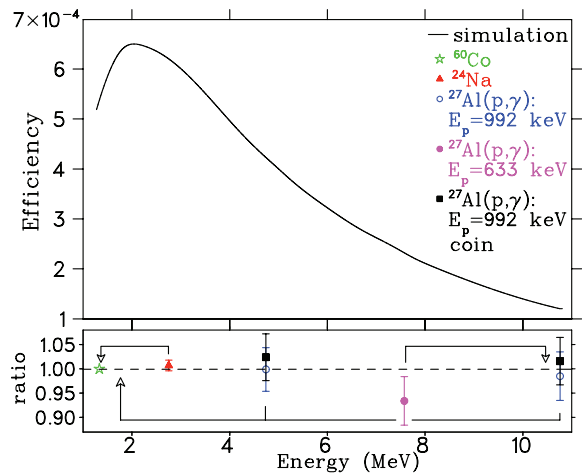}}\hfil
\caption{(color online) Photopeak efficiency.  Top panel is the efficiency from Monte Carlo \penelope\ simulations.  Bottom panel shows the ratio of efficiencies from measurement to simulation, and arrows indicate the gamma-ray energy used in the relative efficiency analysis.  }
\label{eff}
\end{figure}

The gamma rays were projected from their source uniformly in a 80$^\circ$ opening angle, which covered all modeled components, and absolute efficiencies were corrected for the solid angle.  Each simulated energy included an initial number of gamma rays such that the photopeak precision was less than 0.1\%.  At $E_\gamma$ =1332 keV, with the source spread out over an area equal to that of the 1-mm diameter $^{60}$Co source, the simulation initially gave results 2.5\% higher than the measurement.  Therefore, the front face of the crystal was moved back from the target by 1.7 mm, in order to make the simulation reproduce the measurement exactly.  Results for the detector photopeak efficiency are shown in the upper panel of Fig.~\ref{eff}.  The bottom panel of Fig.~\ref{eff} shows the ratio of the measured efficiencies to the simulations.  For sources other than $^{22}$Na, source distribution and substrate material were changed in the simulation to match those used in the measurement.

For the 213- and 610-keV resonances in $^{22}$Na($p,\gamma$), yields from first-escape peaks were added to the photopeak yield in order to improve statistics for branches with $E_\gamma$ = 7333 and 8162 keV, respectively.  Comparison of $^{27}$Al($p,\gamma$) data to simulation at 7.5 MeV indicates agreement to within 2\% and is covered by the systematic error detailed in the next section.  

\subsubsection{Systematic Errors for the Efficiency}\label{effsys}

In order to extract systematic errors for our efficiencies, we compared the quality of the fit of our data from the $^{24}$Na source and the 992- and 633-keV resonances of $^{27}$Al($p,\gamma$) to our simulations, as shown in Fig.~\ref{eff}.  The precisely measured ratio for the two $^{24}$Na gamma rays yields a value for the efficiency at 2754 keV which is in agreement with that given by simulation.  The points obtained from $^{27}$Al($p,\gamma$) resonances have statistical uncertainties between 4.7 and 5.2\%, and they agree well with the simulation.  Therefore we ascribe a 5\% systematic uncertainty to the efficiency determination for isotropic emission of gamma rays.

Because the detectors are centered at $\pm$ 55$^\circ$ in the laboratory, zeros of $P_2(\cos\theta)$, the effect of a $P_2(\cos\theta)$ term in the angular distribution can only arise from the angular dependence of the efficiency across the detector and from center of mass to laboratory transformation.  A $P_4(\cos\theta)$ term has no such restriction.  Assuming the angular distribution to be of the form $\Omega(\theta) = 1 + a_2 P_2(\cos\theta) + a_4 P_4(\cos\theta)$, we used the \penelope\, simulation to determine the effect of non-zero values of $a_2$ and $a_4$.  A value of $a_2$ as large as 1 only caused a 2.6 $\pm$ 0.4\% change in the efficiency.  Published data for $^{23}$Na$(p,\gamma)$ resonances~\cite{glaud} show typical $a_4$ values of about 0.005 and a maximum value of 0.05.  This maximum value would cause a 2.0 $\pm$ 0.4 \% change in the efficiency.  Therefore, we assigned an additional systematic error of 3\%, to include the possible effects of the angular distribution.  Our overall systematic error in the efficiency is $\pm$ 6\%.

\subsection{Beam Properties}

\subsubsection{Rastering}

The beam was rastered using a magnetic steerer located 1 m upstream of the target.   A rectangular pattern with 19- and 43-Hz horizontal and vertical frequencies was used, and signals proportional to the magnetic field were produced by integrating the voltage signals from a pickup coil located in the raster magnet.  These readout values represent the center of the beam spot.  For each data set collected, a 2-dimensional histogram of this signal in both horizontal and vertical directions was recorded.  It was also possible to set a gate on the energy spectrum of each detector and sort out the corresponding raster fields.  
\begin{figure}
\hfil\scalebox{1.1}{\includegraphics{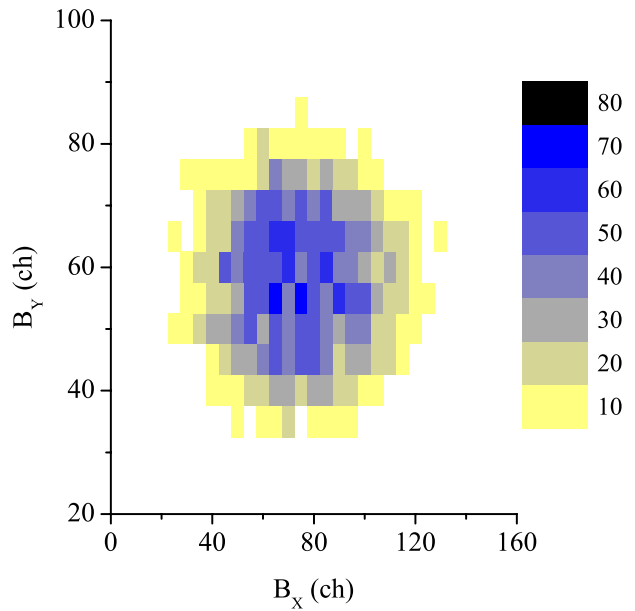}}\hfil
\caption{(color online) 2-dimensional raster plot on the ``coin" target:  a 5-mm diameter $^{27}$Al disc embedded in copper.  B$_{\rm X}$ and B$_{\rm Y}$ are proportional to the magnetic field of the raster.  Shown is a histogram of the raster signal during a measurement of the 406-keV $^{27}$Al($p,\gamma$) resonance, with a wide gate ranging from above the 7358-keV gamma ray to below its first-escape peak.}
\label{coin}
\end{figure}%

Fig.~\ref{coin} shows this two-dimensional raster plot, obtained with a 5-mm diameter, 1.5-mm thick $^{27}$Al disc embedded in the center of a copper backing.  This ``coin" target had the same OFHC copper substrate and diameter as our $^{22}$Na targets.  Shown are the counts detected for each value of the rastering field in the two dimensions.  During all data-taking, the raster signals were monitored in order to diagnose problems with the target or other issues that may have arisen.  For each measured resonance, the amplitude for the raster was scaled as the square root of the proton energy, so the rastered area of the beam on target would be constant.  

\subsubsection{Energy and Density}\label{sim}\label{sys}

The beam energy was determined using a 90$^\circ$ analyzing magnet with an NMR field monitor and calibrated using resonances in $^{27}$Al($p,\gamma$) at $E_p$ = 326.6, 405.5, 504.9, and 506.4 keV.  Given the quality of the fit, we assign a $\pm$ 0.5 keV uncertainty to our knowledge of the beam energy.  


We also conducted measurements on $^{27}$Al targets in order to extract the normalized beam density, averaged over the 5-mm diameter area.  Systematic errors were then explored with simulation and visual inspection of beam-related target coloration.  Beam non-uniformity was then taken into account, and its effect combined with the possibility of a non-uniform target was investigated.

We determined the normalized beam density, $\rho_b$, averaged over the target, by comparing the thick\---target yield from our ``coin" target, $Y_c$, which had the same areal extent as the $^{22}$Na targets, to the yield from a solid $^{27}$Al target, $Y_s$.  From the ratio of these yields, one can extract $\rho_b$ via:

\begin{equation}
\frac{Y_{c}}{Y_{s}} =\rho_b A_c ,\label{beamdensity}
\end{equation}
where $A_{c}$ is the area of the coin.  The yields were measured at different times with different beam tunes using the resonances at $E_p$ = 406 and 992 keV, which yielded results for $\rho_b$ of $2.60 \pm 0.09 {\rm ~cm}^{-2}$ and $2.55 \pm 0.10$ cm$^{-2}$, respectively, with statistical errors only.  The weighted average of these two measurements, 2.58 cm$^{-2}$, was chosen for $\rho_b$.

To study the distribution of beam across the target in order to quantify systematic errors, we carried out a number of measurements using a large raster with $1.8 \times$ the standard amplitude, in addition to the standard amplitude, on the main $^{22}$Na targets and the solid $^{27}$Al and coin targets.  With the raster on and off, transmission measurements through various collimators, shown in Fig.~\ref{chamber}, were performed as well.  These measurements, along with the relative yields of large to standard raster, were used to constrain a Monte Carlo simulation, described below, that was used to investigate potential beam densities on the target.  It also was used to test the effects of possible beam drift, misalignment of the beam and target, and target non-uniformities.  

The simulation modeled transport of the beam through the final components of the beamline and chamber, including the final quadrupole, the rastering unit, and the three sets of collimators.  Variable parameters included beam width and offset, possible collimator offset, raster amplitude, and beam distribution at the quadrupole.  A normalized beam density could not be uniquely determined by this method alone, but the densities within an acceptable phase space, defined by reasonable agreement for transmissions and large/standard rastering yield ratios, ranged from 2.33 to 2.83 cm$^{-2}$.  Even for the extreme case where the quadrupole aperture is filled uniformly, the beam density was found to vary by less than 15\%.  We adopted $\pm$ 0.25 cm$^{-2}$ as the systematic uncertainty in the normalized beam density.

\section{Data and Analysis}

\begin{figure}
\hfil\scalebox{.53}{\includegraphics{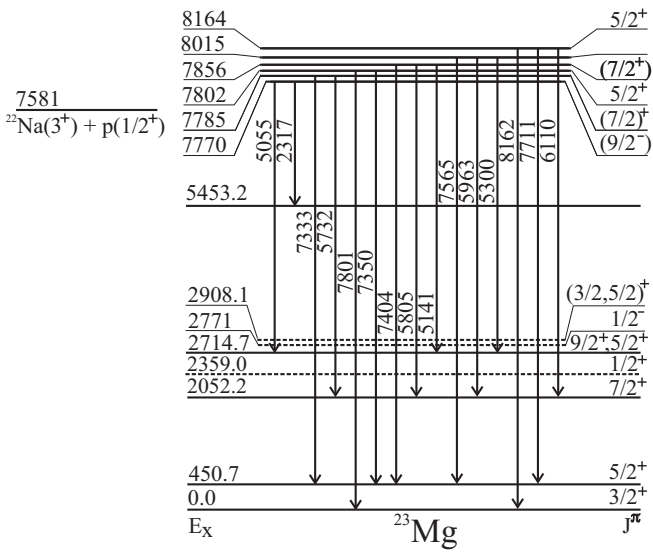}}\hfil
\caption{The relevant part of the decay scheme of $^{23}$Mg.  Energies are in keV.  Not all $^{23}$Mg levels are shown. }
\label{levels}
\end{figure}

Measurements were taken on previously known $^{22}$Na($p,\gamma$) resonances, which we find at $E_p =$ 213, 288, 454, and 610 keV.  We also explored the proposed resonances at 198, 209, and 232 keV.  The relevant energy level diagram for $^{23}$Mg is shown in Fig.~\ref{levels}.  

For all measurements, an energy scan was performed across a $\sim$ 25 keV range about the nominal resonance energy.  To subtract background, which was comprised mostly of cosmic rays and Compton events, we assumed it to have a localized linear dependence on gamma-ray energy and fit the background to windows in the spectrum above and below the line of interest.  This method was especially important for resonances with characteristic gamma rays below 6 MeV.  Here contributions from 6129-keV incident photons due to the contaminating $^{19}$F($p,\alpha\gamma)^{16}$O reaction were significant.  Affected resonances included $E_p$ = 454, 288 keV, and one branch from $E_p$ = 610 keV.  

In order to extract the yield at each laboratory proton energy, $E_p$, an energy window was set on the gamma ray of interest in the vetoed singles spectrum for each of the two germanium detectors.  This window was $\sim$ 25 keV for one detector and $\sim$ 40 keV for the other.  For each detector, the sum of the background subtracted counts in the window is $N_i = N_{\rm prod} \eta_i L_i$, where $\eta_i$ is the efficiency, and $L_i$ is the live time for detector $i$. Then the yield is

\begin{equation}
Y  = \frac{N_{\rm prod}}{Q/e} = \frac{  {N_1}/{L_1}+ {N_2}/{L_2} }{    \eta Q/e } ,
\end{equation}
where $\eta = \eta_1+\eta_2$.  The effects of angular distributions have been addressed in Sec.~\ref{effsys}.  In order to determine the live time, a signal from a pulser unit was fed into the ``test" port of each Ge preamplifier, creating an additional signal in the corresponding amplitude spectra.  This signal was sorted into its own spectrum by a logic signal to the data acquisition.  A window of comparable width to the energy window for the yield was placed on the prompt pulser signal and compared to the scaled number of pulser pulses.  This ratio gave the live time, which ranged from 35 to 45\% for the radioactive targets and was above $\sim$ 90\% for all other targets.  The live-time correction was substantial, but it was not beam-related; instead it resulted from a constant rate due to the radioactivity.  To test the accuracy of the live-time correction, a thick $^{27}$Al target was irradiated with protons with and without a $^{22}$Na source nearby.  Although the presence of the radioactivity decreased the live time from 97\% to 50\%, the ratio of the live-time corrected yields was 0.99 $\pm$ 0.02.

In order to test the sensitivity of our results to the inputs for the gamma-ray background subtraction and its linearity, resonances at $E_p = 610$ keV (with $E_\gamma = 8162$ keV) and 454 keV (with $E_\gamma = 5300$ keV) were inspected, and the choice of window for both the peak and the background on each side was varied to reasonable limits, such as widening, shortening (no window smaller than 10 keV), and shifting.  The resonance strengths changed by less than 1\%, indicating that systematic errors associated with background subtraction are negligible.

After the yields for each excitation function were determined, the areas under the excitation curve were estimated by using the trapezoidal method.  This value for $\int Y_i\, dE$ was then used in Eq.~\ref{main}, along with values determined for all other parameters, to extract the partial resonance strength, $\omega\gamma_i$, for each branch $i$.  The total strength is simply equal to the sum of the partial strengths for all branches.

\subsection{Yields}

\subsubsection{Absolute Yields:  $E_p$ =  610, 454, 288, 213 keV}\label{data}

\begin{figure}
\hfil\scalebox{.4}{\includegraphics{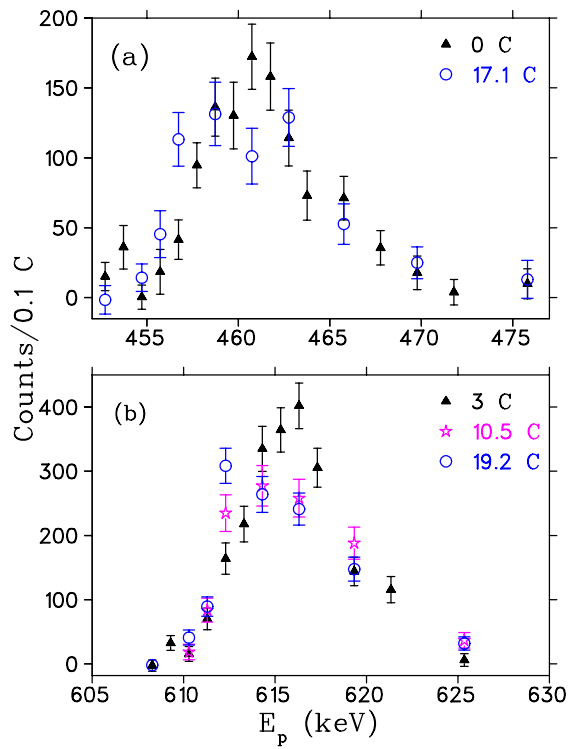}}\hfil
\caption{Excitation functions for (a) $E_p$ = 454 keV and (b) $E_p$ = 610 keV.  $E_p$ is the lab proton energy.  Each plot shows the excitation function at the beginning of target bombardment and at the end, after $\sim$ 20 C had been deposited.  An intermediate curve after 10.5 C is also shown in (b).  (a) is gated on E$_\gamma = 5300$ and 5963 keV, and (b) is gated on the photopeak and single-escape peak of E$_\gamma = 8162$ keV.}
\label{data_613_454}
\end{figure}

\begin{figure}
\hfil\scalebox{.4}{\includegraphics{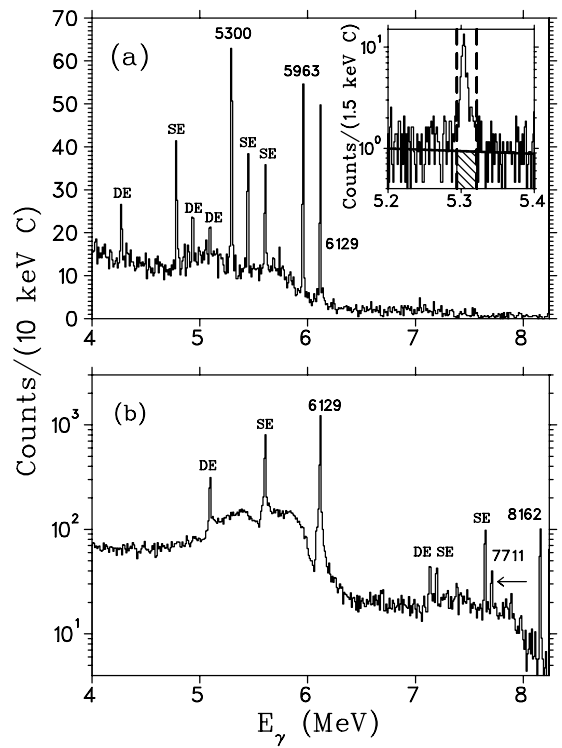}}\hfil
\caption{Summed gamma-ray spectra for (a) $E_p$ = 454 keV and (b) $E_p$ = 610 keV.  The inset illustrates the beam-background subtraction method for E$_\gamma = 5300$ keV, including gamma-ray gate (dashed), analytically calculated background line (solid), and subtracted region (hatched).  SE and DE indicate single- and double-escape peaks, respectively.}
\label{data_s_613_454}
\end{figure}

Fig.~\ref{data_613_454} shows the data taken on the two strongest resonances at $E_p = 454$ and 610 keV.  These resonances were revisited after various amounts of accumulated charge to monitor possible target degradation, discussed in detail in Sec.~\ref{degsys}.  Fig.~\ref{data_s_613_454} shows the corresponding gamma-ray spectra summed over all runs, including an inset illustrating the background subtraction method.  All data for resonances at $E_p$ = 213, 288, and 610 keV were taken on target \#4, and the 454-keV resonance was measured on both targets \#3 and \#4.


\begin{figure}
\hfil\scalebox{1.2}{\includegraphics{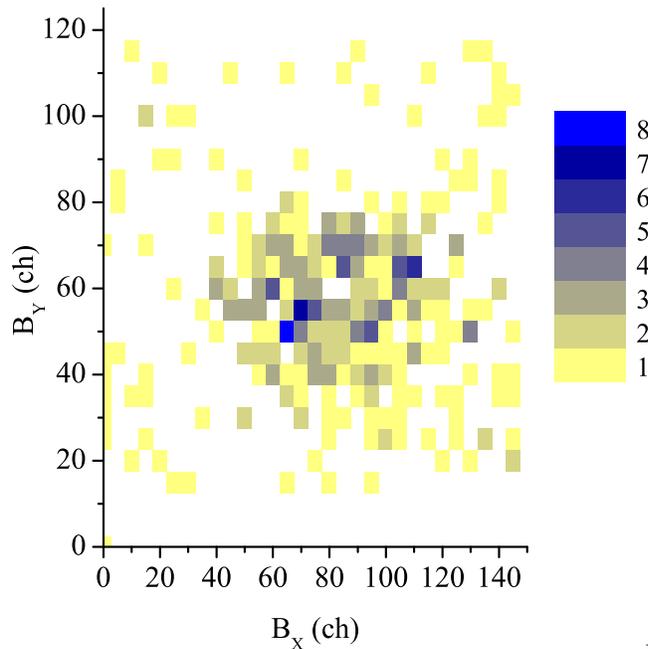}}\hfil

\caption{(color online) Raster plot for the sum of $^{22}$Na($p,\gamma$) for $E_p$ = 454 keV, target \#3, with raster amplitudes $1.8 \times$ larger than our standard amplitudes.  The plot is gated on $E_\gamma$ = 5300 keV.  Outside the target area there are a few bins with only 1 count, which is consistent with the known level of $^{19}$F contamination from the Compton continuum from $E_\gamma$ = 6129 keV. }
\label{raster454}
\end{figure}

Fig.~\ref{raster454} illustrates the summed raster plots for target \#3 with $E_p = 454$ keV.  The concentration of counts from $^{22}$Na($p,\gamma$) are well centered, while a few counts spread through the plot are consistent with yield from $^{19}$F contamination.  

\begin{figure}
\hfil\scalebox{.4}{\includegraphics{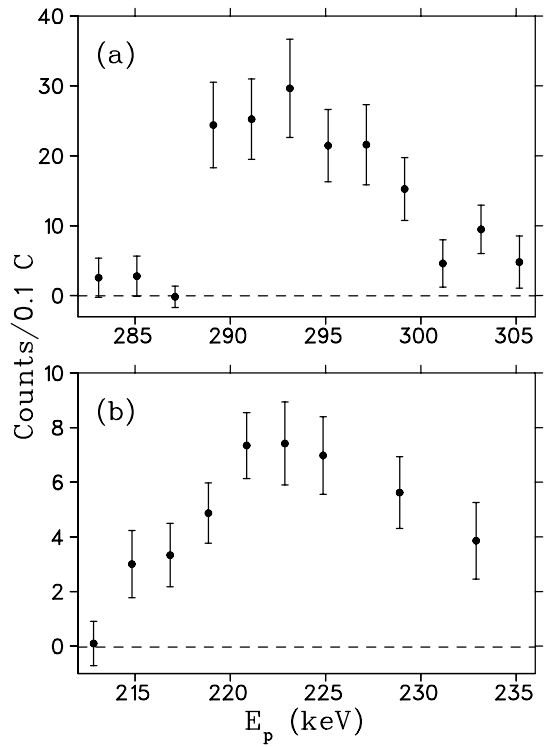}}\hfil
\caption{Excitation functions for (a) $E_p$ = 288 keV and (b) $E_p$ = 213 keV.  $E_p$ is the lab proton energy.  (a) is gated on E$_\gamma = 5141$ keV, and (b) is gated on the photopeak and single-escape peak of E$_\gamma = 7333$ keV.}
\label{data_288_213}
\end{figure}

\begin{figure}
\hfil\scalebox{.4}{\includegraphics{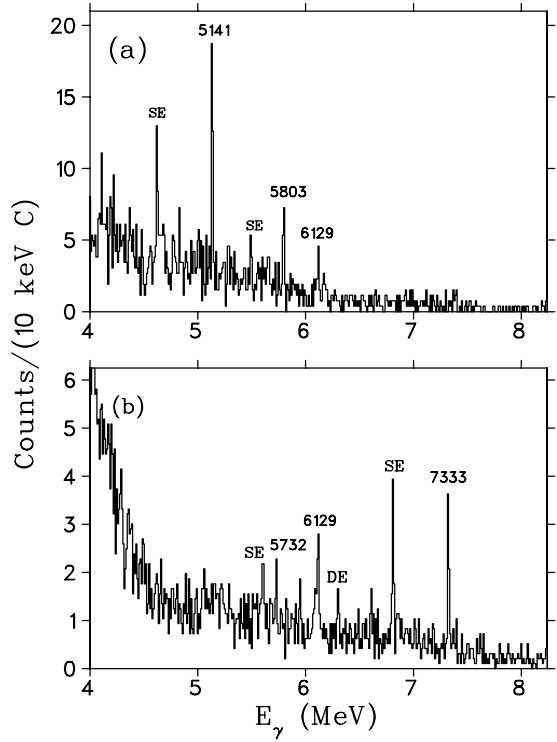}}\hfil
\caption{Summed gamma-ray spectra for (a) $E_p$ = 288 keV and (b) $E_p$ = 213 keV.  SE and DE indicate single- and double-escape peaks, respectively. }
\label{data_s_288_213}
\end{figure}

Fig.~\ref{data_288_213} shows the data taken for 288- and 213-keV resonances, and Fig.~\ref{data_s_288_213} illustrates the summed gamma-ray spectra for each resonance, respectively.  Characteristic gamma rays for each are clearly distinguishable above background.  For $E_p = 213$ keV, the integrated yield was determined with the same trapezoidal method as other resonances, plus a small correction because the highest energy point did not reach zero yield.  The details of determining this contribution are discussed in Sec.~\ref{213}, after a prerequisite analysis method is outlined in the following section.  


\subsubsection{Upper Limits for Yield:  $E_p$ = 198, 232, 209 keV}\label{ULsec}

\begin{figure}
\hfil\scalebox{.4}{\includegraphics{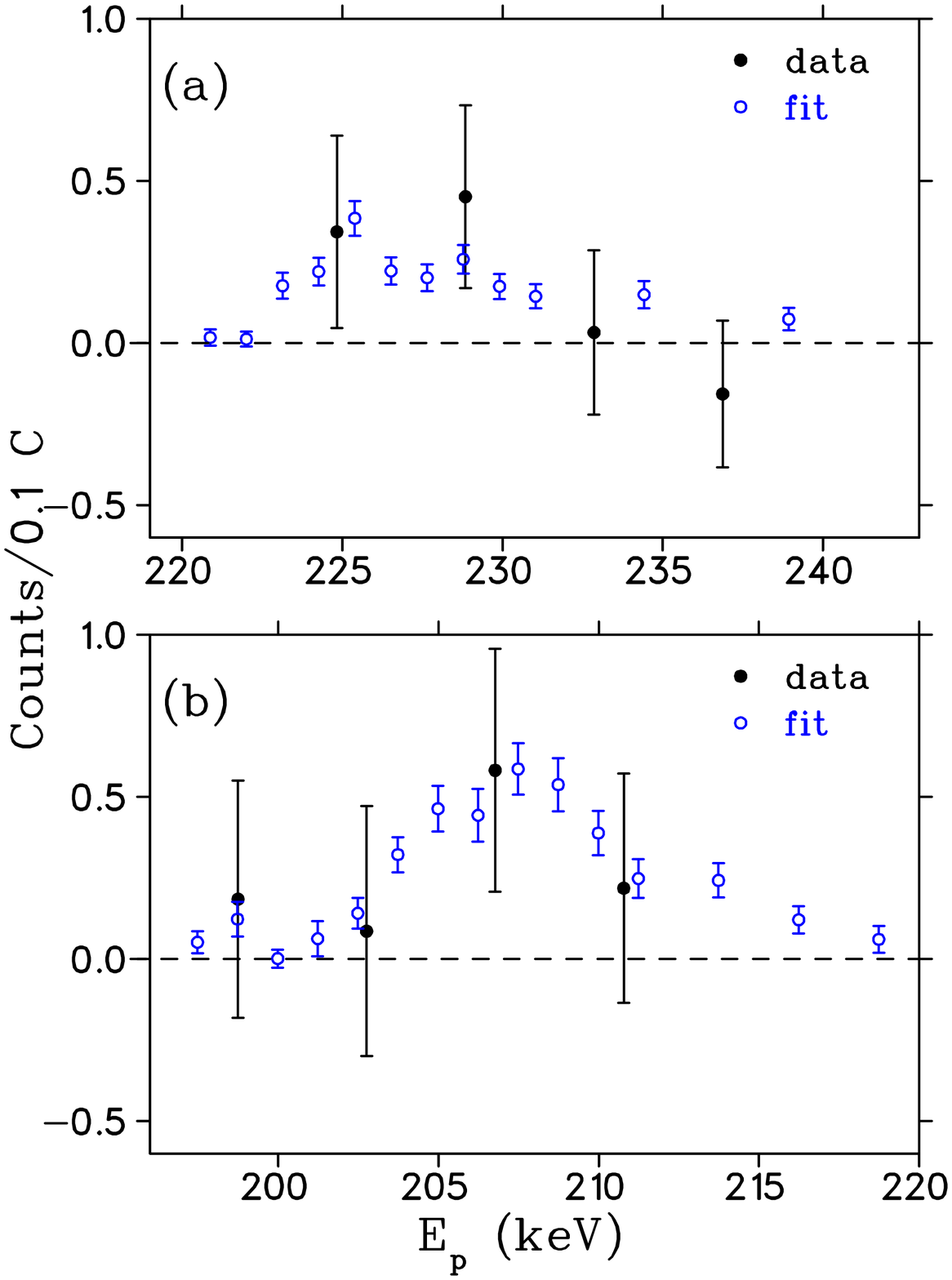}}\hfil
\caption{(color online) Solid circles are excitation functions for (a) $E_p$ = 232 keV and (b) $E_p$ = 198 keV.  $E_p$ is the lab proton energy.  (a) is gated on E$_\gamma = 5055$ keV, and (b) is gated on the sum of E$_\gamma = 7801$ and 7350 keV.  Open circles are the normalized, stretched, and shifted $E_p = 454$ keV excitation function fit, described in the text.  This reference excitation function is not as smooth in (a) as in (b) due to degradation of the bare target.}
\label{data_232_198}
\end{figure}

\begin{figure}
\hfil\scalebox{.4}{\includegraphics{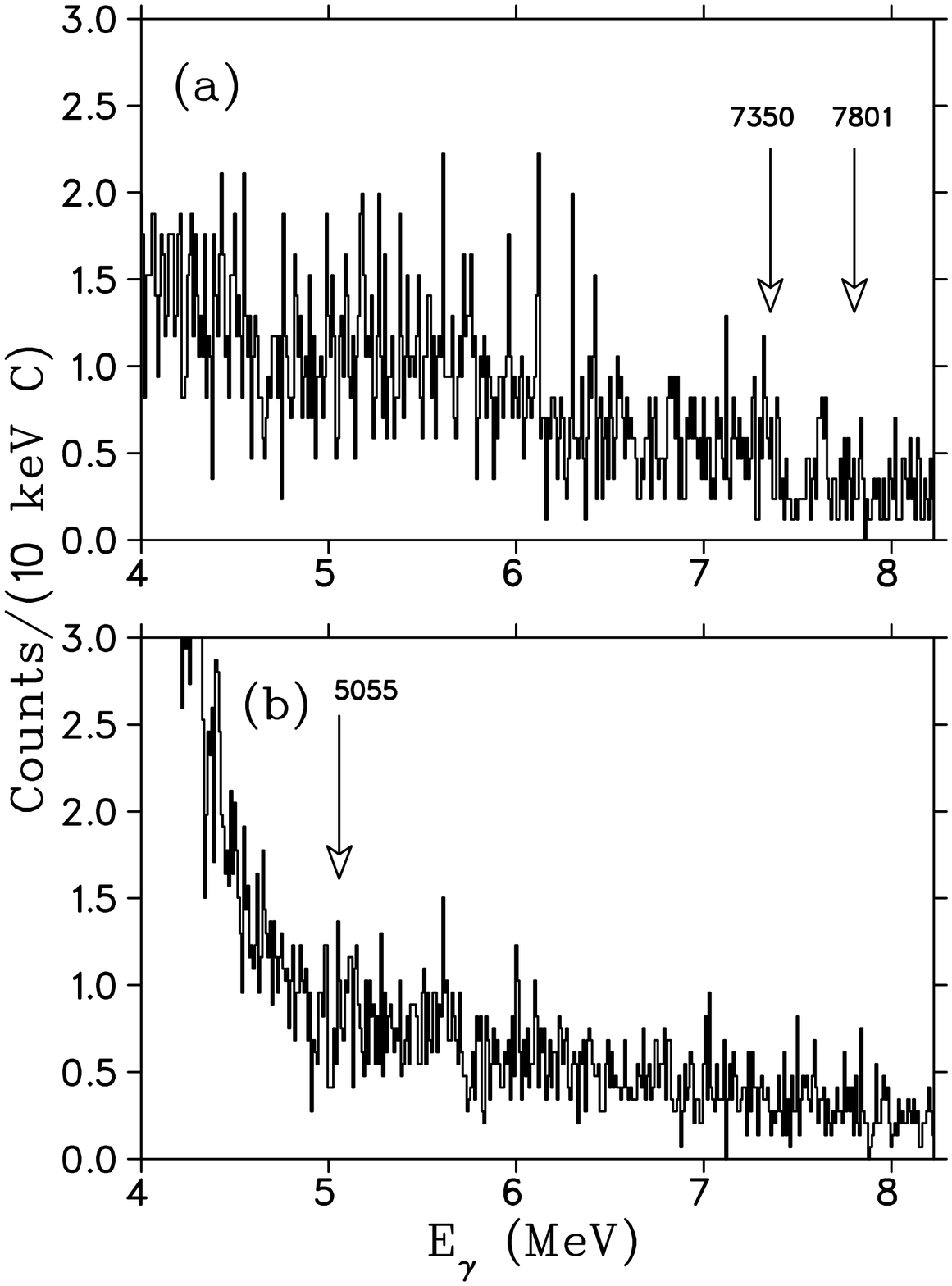}}\hfil
\caption{Summed gamma-ray spectra for (a) $E_p$ = 232 keV and (b) $E_p$ = 198 keV.  The target activity was $2.3\times$ higher in (b) than in (a) and is responsible for the increased background at low energies.  Arrows indicate the energy where one would expect to see gamma rays from the relevant transition.  }
\label{data_s_232_198}
\end{figure}

Data in the region of the proposed resonances at $E_p = 198$ and 232 keV are shown in Fig.~\ref{data_232_198}.  In the summed gamma-ray spectra shown in Fig.~\ref{data_s_232_198}, no discernible gamma-ray yields can be detected above background.  All data for $E_p$ = 198 keV were taken on the chromium-covered target \#3.  Data for $E_p$ = 232 keV were taken on one of the bare test targets, which had previously been exposed to an integrated charge of 13 C.


%

For these resonances, the shape of the excitation function used to determine the area was adopted from either the resonance at $E_p$ = 454 or 610 keV, depending on the target.  Because the resonance shape was dominated by the implantation distribution, this shape was normalized, shifted, and stretched so that it could be fit to the data of the resonance in question.  The stretch factor was fixed and set equal to the ratio of stopping powers in copper for the two energies, whereas the energy shift and the normalization factor were allowed to vary.  The central value of the shift was given by the differences in resonance energies, and the range of the shift was given to fully span the data points.  If a data point for the low-energy resonance fell between points of the normalized curve, the corresponding reference point was determined by a linear interpolation.  For each pair of shift and normalization, the value of the $\chi^2$ between each low-energy resonance and the normalized reference resonance was calculated.  
The modified reference excitation function corresponding to the minimum value of $\chi^2$ is shown for each resonance in Fig.~\ref{data_232_198} (open circles).

\begin{figure}
\hfil\scalebox{.4}{\includegraphics{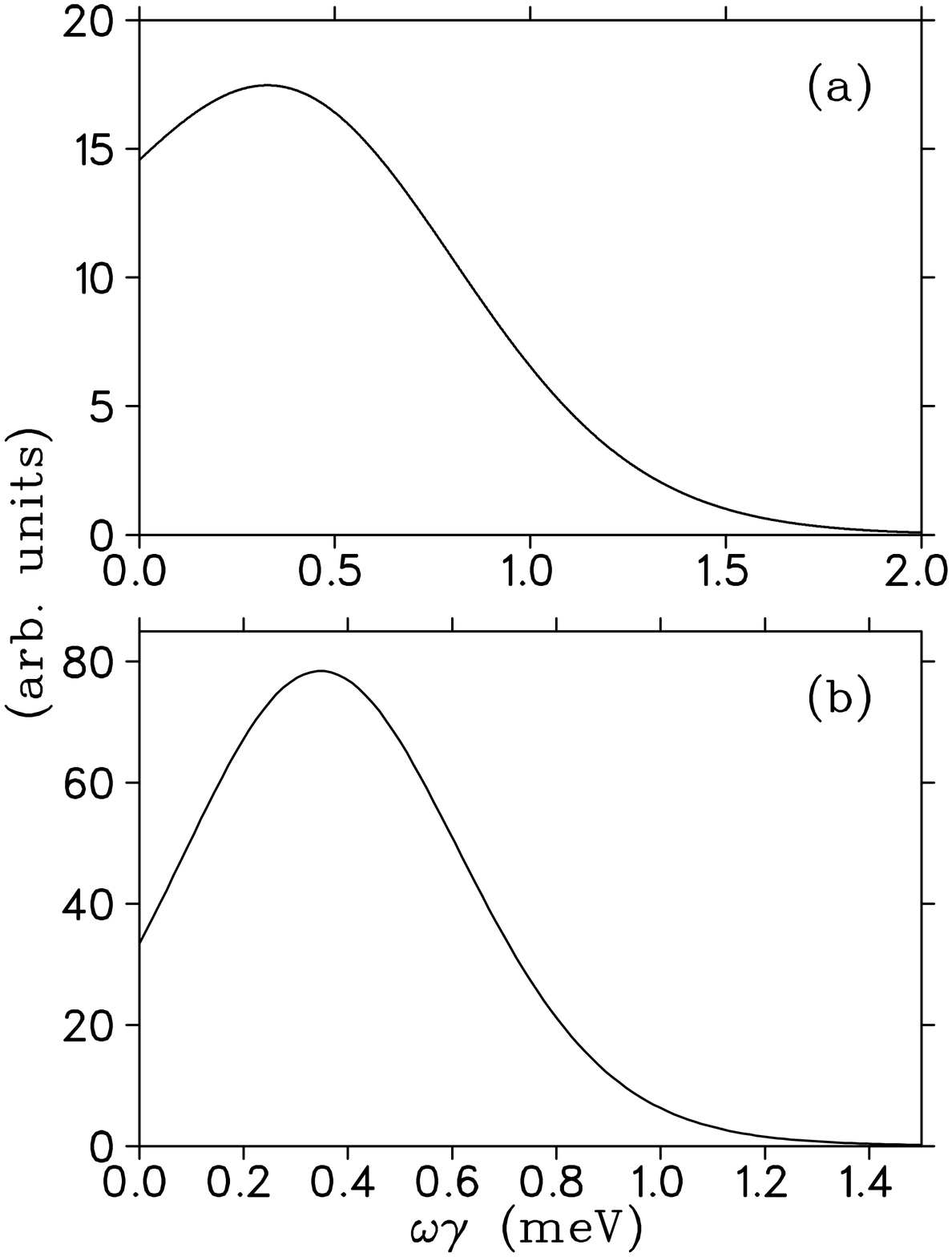}}\hfil
\caption{Projection of exp$(-\chi^2(\omega\gamma_i,E_i)/2)$ onto the $\omega\gamma$-axis for (a) $E_p$ = 232 keV and (b) $E_p$ = 198 keV.  The curves represent the probability density functions, used in Sec.~\ref{astro}.}
\label{pdfs}
\end{figure}%


%

The array of probabilities, $P(\omega\gamma_i,E_i)$, was taken to be proportional to exp$(-\chi^2(\omega\gamma_i,E_i)/2)$, where $\chi^2(\omega \gamma_i, E_i)$ is the $\chi^2$ between the model, assuming particular values for $\omega\gamma_i$ and $E_i$, and the data.  Because we are mainly interested in constraining the value of $\omega\gamma$, we projected the two-dimensional arrays onto the $\omega\gamma$-axis (i.e. $P(\omega\gamma_i)=\sum_{E_i}P(\omega\gamma_i,E_i)$), shown in Fig.~\ref{pdfs}.  The upper limits on $\omega\gamma$ were extracted with a particular confidence level, $C.L.$, using the likelihoods: 

\begin{equation}
C.L. = \frac{\displaystyle\sum_{\omega\gamma_i=0}^{\omega\gamma}P(\omega\gamma_i)}{\displaystyle\sum_{\omega\gamma_i=0}^{\infty} P(\omega\gamma_i)}.
\label{chi2}
\end{equation}
The sum in the denominator was cut off at a maximum value of $\omega\gamma$ such that the sum changed by less than 1\%.  Results are given in Sec.~\ref{results}.

For the possible resonance at 198 keV, a finite value for the strength was also calculated.  Instead of summing from zero and extending upward in the numerator of Eq.~\ref{chi2}, the pair of $\omega\gamma_i$ values with equal values of $P(\omega \gamma_i$) were determined such that the sum between them, properly normalized, gave the desired confidence level.  Because of its small branch, data for the third possible gamma-ray for $E_p = 232$ keV at $E_\gamma$ = 5749 keV was not added to our yield; however, branches from the two other gamma rays were used to adjust the total resonance strength.

For the proposed resonance at $E_p$ = 209 keV, we also applied this method to a hybrid data set comprised of measurements around the resonances at $E_p = 198$ to 213 keV for the gamma ray at 5067 keV, assuming the branch given by Jenkins {\it et al.}~\cite{jenkins} and using the excitation function from the dominant branch of $E_p = 213$ keV and its first-escape peak as the reference curve.  Because the data from the resonance at 198 keV were from a different target, its yields were scaled by the ratio of measured target activities.  The shift in energy was allowed to vary from zero to 25 keV, and the fit yielding the minimum $\chi^2$ was found at the position of the 198-keV data points, possibly due to the fact that the gamma rays have overlapping energy windows.  In other words, we did not observe a separate resonance at $E_p$ = 209 keV.  An upper limit for this resonance, which is presented in Sec.~\ref{results}, was extracted by restricting the energy shift to be equal to the difference in resonance energies, spanning the range of $\pm~ 2\sigma$ around the value claimed by Jenkins $\it{et}$ $\it{al}$~\cite{jenkins}.  


This analysis technique of normalizing and shifting a reference resonance to obtain strengths of others was validated by applying it to the 288-keV resonance, for which the ratio of the strength calculated from this method to the direct method was 0.95 $\pm$ 0.12.

\subsubsection{Corrected Area for $E_p = 213$ keV}\label{213}

In order to estimate the full area of the $E_p = 213$ keV excitation function, a reference resonance measurement at $E_p = 454$ and three at $610$ keV were utilized in the same manner outlined above.  Each of the four curves were fit to the data, and each yielded a data point beyond the fixed $E_p = 213$ keV excitation function that did reach zero.  The last trapezoid area was calculated for each, and the average was added to the area from the direct data, equaling $10 \pm 5$\% of the total area.  The uncertainty in the additional area was set to be the standard deviation among the four fits.   

\subsection{Total Number of Target Atoms}\label{degsys}

\begin{figure}[h]
\hfil\scalebox{.35}{\includegraphics{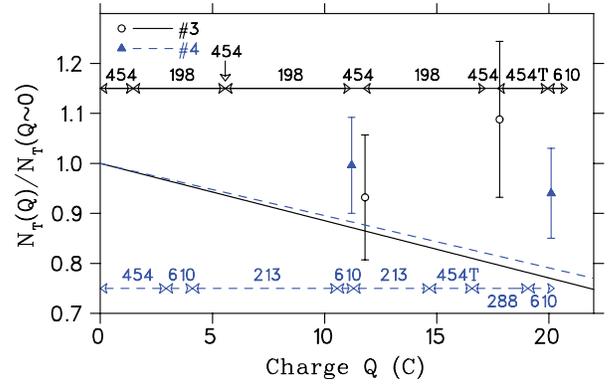}}\hfil
\caption{(color online) Target degradation as a function of accumulated charge for main targets \#3 and \#4.  $N_T(Q)$ is the total number of atoms present in the substrate after an irradiation of charge, $Q$.  The data points illustrate the ratio of the area of each reference excitation function to its initial area at the start of target bombardment.  The lines represent a possible linear decrease of the number of atoms deduced from two residual activity measurements at the beginning and end of bombardment.  Also shown is the timeline for each resonance measurement.  The two sections marked ``454T" indicate variable raster amplitude tests, and although these data are included in the branching ratio determination, they are not included in determining resonance strengths.   }
\label{tdeg}
\end{figure}

We determined the initial number of target atoms from the 1275-keV gamma rays emitted in the decay of $^{22}$Na ($t_{1/2}$ = 2.6027(10) yrs~\cite{nndc}) and assigned a 2.6\% uncertainty.  This uncertainty combines the 1.7\% uncertainty in the $^{60}$Co calibration source at $E_\gamma=1332$ keV with an additional 2\% due to the accuracy of the $\sim$ 7\% background subtraction in this region of high detector rate.  

However, measuring the activity \emph{in-situ} was not sufficient to determine the total number of target atoms throughout the measurement.  During target bombardment, some fraction of $^{22}$Na was sputtered out of the illuminated area of the substrate, yet it remained nearby, maintaining an approximately constant activity throughout the duration of the resonance measurements.  Thus, in addition to determining the total number of initial atoms, monitoring possible target degradation throughout bombardment was particularly important.  

In order to deduce the amount of degradation, two complementary methods were utilized.  One method was to revisit a strong reference resonance periodically throughout the bombardment cycle.  We define $A(Q)$ as the integral of the reference excitation function, $\int Y dE$, after an amount of charge, $Q$ has been deposited.  $A(Q)$ is directly proportional to the number of target atoms, $N_T(Q)$, as shown in Eq.~\ref{main}.  The ratio of the integrals of the excitation function before and after bombardment, $A(Q)/A(Q\sim0)$, is therefore equal to $N_T(Q)/N_T(Q\sim0)$.  The second method, which will be described in detail later in this section, was to measure the residual $^{22}$Na in the chamber before and after target bombardment and use this information to infer the number of sputtered target atoms.  The results of each method are illustrated in Fig.~\ref{tdeg}, along with a timeline for each resonance measurement as a function of accumulated charge. 

Target \#3 accumulated 20.7 C in 186.0 hours, and target \#4 accumulated 20.1 C of charge in 138.3 hours.  Because we covered these targets with 20 nm of chromium, they remained fairly stable.  Shown in Fig.~\ref{data_613_454} (a) are the first and last resonance scans at $E_p$ = 454 keV for target \#3.  Throughout the 20.7 C of bombardment, the resonance at $E_p$ = 454 keV was revisited four times.  $A(Q\sim20$ C$)/A(Q\sim0)$, shown in Fig.~\ref{tdeg}, is 1.07 $\pm$ 0.12, consistent with no target loss.  For target \#4 on which all other non-zero strength resonances were measured, the monitoring resonance was $E_p$ = 610 keV, and multiple scans of its excitation function are shown in Fig.~\ref{data_613_454} (b).  At the end of bombardment, $A(Q\sim20$ C$)/A(C\sim0) = 0.94 \pm$ 0.09, which is consistent with no target loss.  

To use the residual activity method, measurements of the 1275-keV rate were taken before target installation ($R^i_0$), after target installation and before bombardment ($R^i_T$), after bombardment ($R^f_T$), and after target removal ($R^f_0$).  The quantity $(R^i_T-R^i_0) - (R^f_T - R^f_0)$ then is proportional to the amount sputtered from the target.  This final value was used to estimate target degradation throughout the bombardment, assuming linear loss, and is also shown in Fig.~\ref{tdeg} for each target.  As we learned with our $^{23}$Na tests~\cite{tom}, this loss is in fact not linear but usually begins to occur after a significant amount of charge has been deposited, removing the protective layer and sputtering away some of the substrate.  We nevertheless used a hypothesis of linear degradation of the target for one extreme of $N_T$ and an amount consistent with no loss for the other. 

The 198-, 213-, and 232-keV resonance measurements were taken over an extended period of time and charge, whereas all others were measured with a few Coulombs of integrated beam current and did not experience possible prolonged degradation.  The 198- and 213-keV data were taken over $\sim$ 15 and 10 C, respectively, with short interruptions to measure the reference resonance excitation functions.  At the halfway point for each, the linear-decrease hypothesis indicated a 4-5\% loss, although excitation function areas are consistent with no loss at that point.  Therefore, we choose no loss with errors that span the values from each method, $N_T(Q)/N_T(Q\sim0) = 1.00^{+0.00}_{-0.05}$.  Combining this with the systematic uncertainty in the initial number of atoms, we have an overall systematic error of $+2.6$\% and $-5.6$\% in the total number of atoms for the 198- and 213-keV resonance strengths.  For the 454- and 610-keV resonances, which were each measured at the beginning of target bombardment, only an overall systematic error of $\pm 2.6$\% was needed.

The 288-keV resonance was a special case, as its data were not from an extended measurement but were taken after 18 C of irradiation.  Directly following the measurement of this resonance, we performed the final scan of the 610-keV reference resonance, which allowed insight into how many atoms remained.  Therefore, the total number of atoms present for the 288-keV measurement was taken to be the average between linear target loss and loss indicated from the depleted area of the 610-keV resonance curve, $N_T(Q)/N_T(Q\sim0) = 0.88^{+0.12}_{-0.06}$.  Because target loss is not actually linear and could have happened after the 288-keV resonance measurement, the uncertainties span the range between no loss and the value given by linear loss.  

The 232-keV resonance data were taken with a test target with no protective layer after 13 C had already been bombarded, and target loss was appreciable.  At the end of the $\sim$ 20 C irradiation, measurements of the residual activity indicated a 68\% loss.  As explained above, loss is not linear and occurs quite rapidly at the end of the cycle.  Because of this fact, we have chosen to take the loss only from the difference in monitoring the resonance areas, which were taken directly before and after the measurement and gave $A(Q)/A(Q\sim0) = 0.59$.  This value was used to adjust the upper limit on the strength, and we assigned a systematic uncertainty of $\pm$ 40\% to span a wide range approximately down to the value of linear loss.  Regardless, the upper limit on this resonance strength is still dominated by the statistical uncertainty.

\subsection{Systematic Error Summary}

The systematic error budget for resonance strengths is shown in Table~\ref{systab}.  In summary, we find an overall systematic error of $-11.7$\% and $+12.7$\% for the extended measurements at $E_p = 198$ and $213$ keV, resulting from combining uncertainties of $\pm 6$\% in the efficiency, $\pm 10$\% in the normalized beam density, and $-2.6$\% and $+5.6$\% from the number of target atoms.  For the resonances at $E_p = 454$ and $610$ keV, the overall systematic uncertainty is $\pm 11.7\%$, and for the resonance at $E_p = 288$ keV it is $-18.1$\% and $+13.4$\%, differences all due to the differing number of target atoms.  The 232-keV resonance has the largest overall uncertainty, $\pm42\%$.  

\begin{table}
\caption{Summary of systematic errors for each resonance strength.  Note, the total error for the resonances at $E_p = 198$ and 232 keV is dominated by statistical errors.   }
\begin{ruledtabular}
\begin{tabular}{l|cccc}
& \multicolumn{4}{c}{$E_p$ (keV)}\\
Systematic Error & 454/610 & 213/198 & 288 & 232\\
\colrule
Efficiency & 6\%  & 6\%  & 6\%& 6\% \\
Normalized beam density & 10\%& 10\%& 10\%& 10\%\\
Total number of atoms &$ 2.6$\% & $^{+5.6}_ {-2.6}$\% &$^{+7}_ {-14}$\%&$ 40$\% \\
Total & 11.7\% & $^{+12.7}_ {-11.7}$\%  & $^{+13.4}_ {-18.1}$\% &42\%\\
\end{tabular}
\end{ruledtabular}

\label{systab}
\end{table}

\subsection{Resonance Energies}

We were able to obtain resonance energies with two separate techniques.  First, we extracted the resonance energy from the observed gamma-ray energy, along with the excitation energy of the daughter level~\cite{nndc} and Q-value (7580.53 $\pm$ 0.79 keV, using the newly measured masses of $^{23}$Mg~\cite{23mgmass} and $^{22}$Na~\cite{muk}).   From a thick-target $^{27}$Al($p,\gamma)$ measurement at $E_p$ = 406 keV,
the spectrum was calibrated using transitions energies 5088.05 and 7357.84 keV and each first-escape peak from the corresponding gamma rays.  Energies were corrected for Doppler shift and recoil, which ranged in magnitude from $\sim$ 4 to 7 keV and $\sim$ 0.6 to 1.5 keV, respectively.  Because the detector gain depended slightly on rate, a correction ranging in magnitude from 2.4 $\pm$ 0.7 to 3.8 $\pm$ 1.1 keV was also applied.  This was determined by measuring the shift in the calibration gamma-ray energies with and without $^{22}$Na sources nearby.

Second, we found the resonance energy from the proton energy, via the excitation function.  The energy at which the yield reached half its maximum was determined, and the losses in the 20-nm chromium layer and 4 nm of copper were subtracted.  This copper depth is the depth at which the $^{22}$Na distribution reached half of its maximum value.  According to simulations using TRIM~\cite{trim}, the total subtracted losses were $\sim$ 3 to 5 keV, and a 20\% uncertainty in the stopping power was assumed.  For the 213- and 288-keV resonances, an additional adjustment to the resonance energy was added to account for the slight transformation of the excitation functions due to sputtering, as shown in Fig.~\ref{data_613_454} (b).  From repeated scans of the 610-keV resonance, the energy at half of the maximum yield changed by 1.2 $\pm$ 0.9 keV after 11 C, in the middle of the 213-keV resonance measurement, and that shift remained constant after 19 C, directly after the 288-keV measurement.  Those resonance energies were adjusted by that amount.  Results are given in Sec.~\ref{results}.  Similarly, excitation energies, $E_x$, and gamma-ray energies, $E_\gamma$, each were found independently by using the weighted average of the respective value extracted from the excitation function and the respective value extracted from the gamma-ray spectra.    

\subsection{Branches}\label{brA}

Strong branches were determined from the spectra summed over all runs within a particular resonance.  For resonances at $E_p = 454$ and 610 keV, which were used as reference resonances to monitor degradation and for other target tests, the total amount of data was significantly larger than for a single resonance scan.  

Due to the very low statistics for weaker possible branches, an additional restriction was placed on the analysis.  Similar to our analysis for weak resonances where a reference resonance excitation function was shifted and normalized to fit the data, the excitation function for the strongest branch (and its first-escape peak for branches from $E_p = 610$ keV) was normalized to match the weaker branches' excitation functions, such that the $\chi^2$ was minimized.  Then the branches were extracted using Eq.~\ref{chi2}.  

Only an upper limit for the branch at $E_\gamma = 6112$ keV from the $E_p = 610$ keV resonance could be obtained due to the obscuring peak at $E_\gamma = 6129$ keV from $^{19}$F contamination.  In order to estimate the contribution from the $^{22}$Na($p,\gamma$) resonance, the spectral line shape was deconvolved into the contribution from $^{19}$F and from $^{22}$Na.  This was done by comparing the on-resonance line shape with the sum of the off-resonance line shape and a shape representing the 6112-keV gamma ray.  The latter shape was estimated from a normalized and shifted peak at $E_\gamma = 8162$ keV, the largest branch in the de-excitation.  The value of this normalization was equal to the ratio of the magnitude of the branches, adjusted for efficiency differences, and was extracted by minimizing the $\chi^2$ of the hybrid curve and the on-resonance curve.  Because of the changing shape of the off-resonance curve above and below the resonance, only an upper limit could be obtained.

\subsection{Verification of Experimental Method}

\begin{figure}
\hfil\scalebox{.35}{\includegraphics{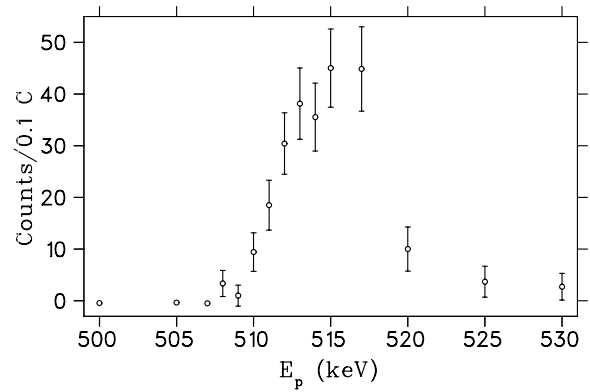}}\hfil
\caption{Excitation function for $^{23}$Na($p,\gamma$) resonance at $E_p$ = 512 keV gated on $E_\gamma = 10810$ keV.  $E_p$ is the lab proton energy.}
\label{512}
\end{figure}
In order to verify our technique, targets of $^{23}$Na were implanted using the ion source on the injector deck at the low-energy end of the University of Washington accelerator, and known $^{23}$Na($p,\gamma$) resonances were measured under the same conditions as the $^{22}$Na($p,\gamma$) measurements.  A 20-nm layer of chromium was also evaporated on the surface, similar to the main $^{22}$Na targets.

The $^{23}$Na($p,\gamma$) resonance at $E_p$ = 512 keV has a reported strength of 91.3 $\pm$ 12.5 meV and is the recommended reference resonance for this reaction~\cite{iliadis}.  We measured this with a target containing (5.8 $\pm$ 0.9) $\times 10^{15}$ atoms, as determined by implantation charge integration with a measured correction for the sputtering of positive ions.  The excitation function is shown in Fig.~\ref{512}. 
Using the 10810-keV gamma ray with a branch of 71\%~\cite{23nabranch}, we determined its total resonance strength to be 79 $\pm$ 17 meV, after applying the correction described below.

Inspection of the raster plot indicated a small part of the target was missed when using our standard raster.  To determine the missing fraction, the target was illuminated with a larger raster amplitude that fully encompassed the target atoms.  Comparing the full target area to the fractional area covered with the standard raster, the amount missed was estimated to be $\sim$ 12\%.  This correction was applied to the resonance strength.  This was not necessary with our $^{22}$Na targets, as their respective raster plots indicated the beam covered the entire active area.  To estimate a range for the effect of the observed $^{23}$Na target misalignment, simulations were performed, as described in Sec.~\ref{sim}, with an offset of 1.5 to 2.0 mm and a variety of target distributions.  The uncertainties associated with these calculations dominate the overall uncertainty quoted above.

\section{Results and Discussion}\label{results}

\begin{table*}
\centering
\caption{Resonance energies.  Values are extracted from the lab proton energy, $E_p$, via the excitation function, correcting for energy loss $E_{\rm loss}$, and from the gamma-ray energy, $E_\gamma$.  Statistical uncertainties are given for each individual branch, and combined branches include systematic uncertainties.  The adopted value is the weighted average of both methods.  All values are in keV.}
\begin{ruledtabular}
\begin{tabular}{c|c|ccc|cc|c}
Previous & \multicolumn{7}{c}{Present}\\
\hline
&&&\multicolumn{2}{c|}{$E_p$ (from excitation function):}&\multicolumn{2}{c|}{$E_p$ (from $E_\gamma$):}&\\
$E_p$\footnotemark[1] & E$_\gamma$&$E_{\rm loss}$ & branch& combined branches  & branch &combined branches& Adopted $E_p$ \\
\hline
213.3 $\pm$ 2.7 &7332.7 $\pm$ 1.2 &  5.1 $\pm$ 1.0 &213.1 $\pm$ 3.0 &213.1 $\pm$ 3.4&213.6 $\pm$ 1.6& 213.6 $\pm$ 1.6&213.5 $\pm$ 1.4\\
\hline
287.9 $\pm$ 2.1 &5140.6 $\pm$ 1.0 &  4.6 $\pm$ 0.9&286.5 $\pm$ 2.1 & 286.3 $\pm$ 2.3&288.4 $\pm$ 1.7&288.7 $\pm$ 1.3 &288.1 $\pm$ 1.1\\
& 5803.2 $\pm$ 1.3&&285.6 $\pm$ 4.2&& 289.1 $\pm$ 2.0&&\\
\hline
457 $\pm$ 2\footnotemark[2] &5300.1 $\pm$ 0.8& 3.7 $\pm$ 0.8&452.8 $\pm$ 0.8&452.8 $\pm$ 1.1& 455.6 $\pm$ 1.7& 455.7 $\pm$ 1.1&454.2 $\pm$ 0.8\\
&5962.7 $\pm$ 0.8 &&452.8 $\pm$ 1.1   & & 455.9 $\pm$ 1.6&&\\
\hline
611.3 $\pm$ 1.8 &8162.3 $\pm$ 0.9& 3.2 $\pm$ 0.7& 609.6 $\pm$ 0.9 &609.0 $\pm$ 1.1 & 611.0 $\pm$ 1.5&610.8 $\pm$ 1.2&609.8 $\pm$ 0.8\\ 
&7711.2 $\pm$ 1.1 &&608.1 $\pm$ 1.1 && 610.3 $\pm$ 2.0&&\\
\end{tabular}
\end{ruledtabular}
\footnotetext[1]{From Ref.~\cite{stegmuller}, unless otherwise noted.}
\footnotetext[2]{From Ref.~\cite{seuthe}.}
\label{tab1}
\end{table*}

Results are shown in Tables~\ref{tab1},~\ref{br}, and~\ref{table}.  The resonance energies are summarized in Table~\ref{tab1}, the gamma-ray branches and partial strengths for each resonance are summarized in Table~\ref{br}, and the final resonance strengths are summarized in Table~\ref{table}.

The resonance energies determined from each method are shown in Table~\ref{tab1}, and agreement between methods is quite good.  The adopted energy is the weighted average of the two results.  We find energies that agree with previously reported values~\cite{seuthe,stegmuller,jenkins}, and we have improved the uncertainties on the energies.

\begin{table*}
\caption{Branches and partial strengths of $^{22}$Na($p,\gamma$).  $E_x$ and $E_t$ are the excitation and transition energies, respectively.  The value given for the branch was determined from the sum of all data for a particular resonance.  Systematic errors are not included in the partial strengths.  68\% confidence levels are given for all data, and both an upper limit and measurement have been given for $E_p = 198$ keV.  Total strengths are shown in Table~\ref{table}.  All upper limits for partial strengths were derived assuming their excitation function was the same shape as for the strongest branch. }
\begin{ruledtabular}
\begin{tabular}{lcccc|ccccc|c}
&&&&&\multicolumn{5}{c|}{Branches (\%)\footnotemark[1]}&\\
$E_x$\footnotemark[2] &$E_p$ & $E_t$\footnotemark[2]   &$E_f$&&\multicolumn{4}{c}{Previous\footnotemark[3]}&Present&$\omega\gamma_{\rm partial}$ \\
(keV) &(keV)&(keV)& (keV)&I$^\pi_f$&Ref.~\cite{seuthe}&Ref.~\cite{stegmuller}\footnotemark[4]&Ref.~\cite{jenkins}&Ref.~\cite{iacob}\footnotemark[4] & & (meV)\\
\colrule

7770.2$\pm$1.4&198 & 5055&2715&9/2$^+,5/2^+$& -&-&58$\pm$8&-& - &$\leq 0.30$ (0.20$^{+0.15}_{-0.13}$) \\
&& 2317&5453&-&- &-&42$\pm$7&-& - &-\footnotemark[5]\\

\hline
7782.2$\pm$1.2&209&5730 &2052&7/2$^+$&-&& 33$\pm$6&-& -& -\footnotemark[6]\\
&& 5067 &2715&9/2$^+,5/2^+$&-&-&66$\pm$8&-& -  & $\leq 0.26$ \\
\hline

7784.7$\pm$1.2&213& 7785&0& 3/2$^+$&-&$\leq 29$&-& 3.1$\pm$2.0 & -  & $\leq 0.09$\\
&&7334&451&5/2$^+$&-&100 &100&80.8$\pm$3.6& 89.4$\pm$5.3& $5.1 \pm 0.5$  \\
&& 5732&2052 &7/2$^+$&-&$\leq 29$&-& 16.2$\pm$3.4& 10.6$\pm$5.3\footnotemark[7]  & 0.6$\pm$0.3\footnotemark[7]\\
&& 5426&2359 &1/2$^+$&-&$\leq 29$&-& - & -  & $\leq 0.24$ \\
&& 5070&2715 &$9/2^+,5/2^+$&-&$\leq 29$&-& - & -  & $\leq 0.33$ \\
&& 5014 &2771&$1/2^-$&-&-&-& - & -  & $\leq 0.17$\\
&& 4877&2908 &$(3/2,5/2)^+$&-&-&-&- & -  & $\leq 0.24$ \\

\hline
7802.2$\pm$1.4&232& 7802&0&3/2$^+$&-&-&-& 66.4$\pm$2.4& -&$\leq 0.44$\\
&& 7351 &451&5/2$^+$&-&-&-& 29.9$\pm$2.4 & -  & $\leq 1.04$\\
&& 5750 &2052&7/2$^+$&-&-&-& 3.7$\pm$1.3 & -  &-\\

\hline
7856.1$\pm$1.0&288 & 7856 &0&3/2$^+$&-&$\leq 4.3$ &-& - & -  & $\leq 0.69$\\
&& 7405 &451&5/2$^+$&-& 10.8$\pm$2.9&-&-& 6.7$\pm$2.9 & 2.6$\pm$1.5\\
&& 5804 &2052&7/2$^+$&36$\pm$12& 27.2$\pm$2.7&-&-& 26.2$\pm$4.1&  10.4$\pm$1.9\\
&& 5497 &2359&1/2$^+$ &-&$\leq 4.3$&-&-&-& 2.2$\pm$1.6\footnotemark[8]\\
&& 5141&2715 &9/2$^+$,5/2$^+$&64$\pm$12& 62.0$\pm$3.1&100&-&67.1$\pm$4.4 & 26.2$\pm$2.9\\
&& 5085 &2771&1/2$^-$ &-&- & -  &-&-& $\leq 0.52$\\
&& 4948&2908 &$(3/2,5/2)^+$&-&- & -&-& - & $\leq 0.81$\\

\hline
8015.3$\pm$0.8&454&8015 &0&3/2$^+$&-&-&-& - & -& -\footnotemark[9]\\
&&7565 &451&5/2$^+$&-&-&-& - & 4.5$\pm$0.8 & 5.3$\pm$1.7\\
&& 5963 &2052&7/2$^+$&-&-& 29$\pm$12&-& 43.6$\pm$1.2& 74.7$\pm$5.2\\
&& 5656 &2359&1/2$^+$&-&-&-& - & -  & $\leq 1.1$ \\
&&5301&2715&$9/2^+,5/2^+$& 100&-&71$\pm$16&-&51.9$\pm$1.2& 85.9$\pm$5.5 \\ 
&& 5244&2771 &$1/2^-$&-&-& - &-& -  & $\leq 0.8$ \\
&& 5107 &2908&$(3/2,5/2)^+$&-&-& - &-&-  & -\footnotemark[10]\\
\hline

8163.9$\pm$0.8&610& 8164 &0&3/2$^+$& 65$\pm$5&65.0$\pm$2.3&-&-& 61.3$\pm$1.8& 376$\pm$16\\
&& 7713&451 &5/2$^+$&19$\pm$2& 15.0$\pm$1.8 &-&-& 18.6 $\pm$1.3& 97$\pm$16\\
&& 6112 &2052&7/2$^+$&16$\pm$2& 20.0$\pm$1.8 &100&-& (20.0$\pm$1.8)\footnotemark[11]&  (118$\pm$13)\footnotemark[12] \\
&& 5805 &2359&1/2$^+$&-&$\leq 1.4$ & - &-&-& $\leq 24$ \\
&& 5449&2715 &9/2$^+$,5/2$^+$&-&$\leq 1.4$ &-&-& -  & $\leq 11$ \\
&& 5393 &2771&1/2$^-$&-&- & -  &-&-&$\leq 18$\\
&& 5256 &2908&$(3/2,5/2)^+$&-&- & -&-& - & $\leq 18$ \\
\end{tabular}
\end{ruledtabular}
\footnotetext[1]{We assume the sum of all observed branches adds up to 100\%.}
\footnotetext[2]{Derived from our results in Table~\ref{tab1}.  Otherwise from NNDC~\cite{nndc}. }
\footnotetext[3] {Branches from upper limits are calculated as the partial strength relative to the total observed strength.  }
\footnotetext[4]{Converted finite values into percents.}
\footnotetext[5]{Partial strengths cannot be determined.  See Table~\ref{table} for upper limits on total strengths.}
\footnotetext[6]{Any contribution from this branch has been attributed to $E_p = 213$ keV.}
\footnotetext[7]{Imposed the restriction that the shape of the excitation function must be the same as for the strongest branch.}
\footnotetext[8]{Because I$_f$ = 1/2, this state is highly unlikely to have a detectable value so we attribute this value to statistical fluctuations and do not include it in $\omega\gamma_{\rm total}$. }
\footnotetext[9]{Could not be determined due to a small percentage of pulser counts sorted into this energy window. }
\footnotetext[10]{Could not be determined due to $^{19}$F background. }
\footnotetext[11]{The value from Ref.~\cite{stegmuller} is used, as this gamma ray was obscured by $^{19}$F background in our measurements.}
\footnotetext[12]{Estimated from branch.}
\label{br}
\end{table*}

\begin{table}
\caption{$^{22}$Na($p,\gamma$) resonance strengths.  Upper limits are at the 68\% confidence level.  Both an upper limit and measurement have been given for $E_p = 198$ keV.  The total strength was equal to the sum of partial strengths given in Table~\ref{br} for all resonances, with the exception of $E_p$ = 198, 209, and 232 keV where branches from Refs.~\cite{jenkins,iacob} must be used.  If only an upper limit has been given to a particular branch in Table~\ref{br}, then its value directly increased our upper uncertainty only, not the central value. }
\begin{ruledtabular}
\begin{tabular}{llcc}
&Present & Previous & Present\\
$E_p^{\rm lab}$&$E_p^{\rm cm}$ (keV)& $\omega\gamma_{\rm total}$ (meV)\footnotemark[1] & $\omega\gamma_{\rm total}$ (meV) \\
\colrule
45&43.1 $\pm$ 1.7\footnotemark[2] &(7.1 $\pm$ 2.9)$\times 10^{-14}$\footnotemark[2] &(7.1 $\pm$ 2.9)$\times 10^{-14}$\footnotemark[2]  \\
70&66.6 $\pm$ 3.0\footnotemark[2]  &(5.1 $\pm$ 2.1)$\times 10^{-10}$\footnotemark[2] &(5.1 $\pm$ 2.1)$\times 10^{-10}$\footnotemark[2] \\
198&189.5 $\pm$ 1.8\footnotemark[3]   & $\leq 4$\footnotemark[3]  & $\leq$ 0.51 (0.34$^{+0.25}_{-0.22} $ ) \\
209&200.2 $\pm$ 1.6\footnotemark[3]   &$(5 \times 10^{-2})$\footnotemark[3]  & $\leq$ 0.40\footnotemark[4]\\
213&204.1 $\pm$ 1.4  & 1.8 $\pm$ 0.7\footnotemark[5] & $5.7^{+1.6}_{-0.9} $ \\
232&221.4 $\pm$ 2.3\footnotemark[3]   &2.2 $\pm$ 1.0\footnotemark[3]  & $\leq$ 0.67\\
288&275.4 $\pm$ 1.1 & 15.8 $\pm$ 3.4 & 39 $\pm$ 8\\
454&434.3 $\pm$ 0.8  & 68 $\pm$ 20 & $166 \pm 22$\\ 
503&481 $\pm$ 2\footnotemark[1] & 37 $\pm$ 12& 93 $\pm$ 36\footnotemark[6] \\
610&583.1 $\pm$ 0.8 & 235 $\pm$ 33 & $591^{+103}_{-74}$\footnotemark[7]\\
740&708 $\pm$ 2\footnotemark[1] & 364 $\pm$ 60& 913 $\pm$ 174\footnotemark[6] \\
796&761 $\pm$ 2\footnotemark[1] & 95 $\pm$ 30& 238 $\pm$ 79\footnotemark[6] \\

\end{tabular}
\end{ruledtabular}
\footnotetext[1]{From Ref.~\cite{seuthe,stegmuller}.}
\footnotetext[2]{From or derived from Ref.~\cite{iliadispreprint3}.}
\footnotetext[3]{From Ref.~\cite{jenkins}.}
\footnotetext[4]{Not included in the present reaction rate.}
\footnotetext[5]{Actual value measured was 1.4 meV, inflated to 1.8 meV to account for possible unknown branches.}
\footnotetext[6]{Scaled from Ref~\cite{seuthe}.  Not measured in this work.}
\footnotetext[7]{Includes estimated partial branch from $E_\gamma = 6112$ keV.}
\label{table}
\end{table}

Table~\ref{br} shows our excitation energies, gamma-ray energies, branches, and partial strengths.  Also included is a comparison with previous branches.  Our branches are in agreement with the previous direct measurements~\cite{seuthe,stegmuller} for $E_p = 288$ and 610 keV.  For the 6112-keV branch of the 610-keV resonance that could not be resolved from the $^{19}$F contamination, we found the upper limit to be 28\%, which is consistent with the value of 20.0 $\pm$ 1.8\% measured directly by Stegmuller {\it et al.}~\cite{stegmuller}.  We have adopted the branch of Ref~\cite{stegmuller} to extract its partial strength because it is consistent with ours and is more precise.  For the previously established branches of the 454-keV resonance, our branches agree with Jenkins {\it et al.}~\cite{jenkins}, and we have improved upon their uncertainties.  A new branch has also been identified with $E_\gamma = 7566$ keV.  Although branches for the 213-keV resonance are in agreement with the earlier measurement~\cite{stegmuller} at the $2\sigma$ level, an additional branch of $\sim 11\%$ has since been identified, and we agree with its currently established value~\cite{iacob}.  

For the observed resonances at $E_p = 213$, 288, 454, and 610 keV, contributions to the total strength have been investigated for primary transitions to all levels up to the 6th excited state, also shown in Table~\ref{br}.  For unobserved branches, limits were obtained via the method described in Sec.~\ref{brA}.  The branches with a final spin of 1/2 are highly unlikely to be detectable based on angular momentum considerations, so we attribute our non-zero result for $E_\gamma = 5497$ keV from the 288-keV resonance to statistical fluctuations.  Upper limits on partial strengths from spin 1/2 levels have not been included in the strength uncertainties or in the branches. Other branches where only an upper limit could be determined were used to directly increase the upper uncertainty on the total strength but do not affect the central value of the total resonance strength.  

All finite total strengths are significantly larger than those previously reported~\cite{seuthe,stegmuller}, as shown in Table~\ref{table}.  The 213-keV resonance is stronger by a factor of 3.2, which includes a factor of 2.8 that we observe with respect to the same decay channels observed in Ref.~\cite{stegmuller}.

Iacob {\it et al.}~\cite{iacob} attempted to extract a resonance strength for $E_p = 213$ keV by compiling data from several different sources.  Using the beta-delayed proton branch from Tighe {\it et al.}~\cite{tighe}, the proton-to-gamma-ray branching of Per{\"a}j{\"a}rvi {\it et al.}~\cite{per}, and the lifetime of Jenkins {\it et al.}~\cite{jenkins}, Iacob {\it et al.} claimed a total strength of 2.6 $\pm$ 0.9 meV, in contrast to our value of $5.7^{+1.6}_{-0.9}$ meV.  We consider the value of Iacob {\it et al.} to be unreliable.  It is based on the observation of a $\beta$-delayed proton peak very close to detection threshold that would have been difficult to disentangle from noise.  Indeed, Saastamoinen {\it et al.}~\cite{saas} have shown that the $\beta$-delayed proton intensity deduced by Tighe {\it et al.}~\cite{tighe} was too large by a significant amount. In addition, Iacob {\it et al.}'s value is based on the lifetime of the state, for which further documentation has not been published.

We have set upper limits on the 198- and 232-keV potential resonances.  For $E_p = 198$ keV, based on a search for the 5055-keV branch, we observe a possible presence in our fits with a non-zero strength at slightly over $1\sigma$.  Jenkins {\it et al.}~\cite{jenkins} suggested that this resonance could have a strength as high as 4 meV, whereas our 68\% $C.L.$ upper limit is a factor of 8 smaller.  The resonance at $E_p = 232$ keV also was not observed, based on our search for the two main branches reported by Ref.~\cite{jenkins}.  This resonance has been demonstrated to be the isospin analogue of the $T=3/2$ $^{23}$Al ground state~\cite{saas,iacob} so its width for proton decay to $T=0$ $^{22}$Na is significantly suppressed by isospin conservation, and thus it is expected to have a relatively small $(p,\gamma)$ resonance strength.  Our direct measurement is consistent with this expectation.

Jenkins {\it et al.}~\cite{jenkins} proposed a new level that corresponds to $E_p$ = 209.4(17) keV, which produces gamma rays with energies of 5729.1(11) and 5067.1(11) keV and branches of 33(6)\% and 66(8)\%, respectively.  The gamma ray at 5067 keV is very close in energy to the 5055-keV gamma ray from the possible resonance at 198 keV, as they produce the same final state.  We have investigated this resonance, but due to the size of the energy windows necessary on each peak ($\sim$ 20 to 45 keV), it is unclear from which potential resonance these possible gamma rays originate.  Contrary to Jenkins {\it et al.}, Iacob {\it et al.}~\cite{iacob} attributed the gamma ray at 5729 keV to the resonance at 213 keV.  Our branching ratios in Table~\ref{br} are consistent with Iacob {\it et al.} and not with Jenkins {\it et al.}  An upper limit of 0.40 meV can be placed on the strength of the 209-keV resonance at the 68\% confidence level, and it is possible that part of the contribution originates from the potential resonance at 198 keV.

Seuthe {\it et al.}~\cite{seuthe} also measured resonances at $E_p = 503, 740,$ and 796 keV that we did not investigate.  These resonances do not play an important role in the nova scenarios we consider here, but we nevertheless included our estimate for their contributions in our calculation of the thermonuclear reaction rate.  We assumed that the {\it relative} strength of the resonances to the 454- and 610-keV resonances is correctly given by the result of Seuthe {\it et al.}  Because we observe an average factor of 2.5 $\pm$ 0.5 larger for the strengths of resonances observed here with respect to those in Ref.~\cite{seuthe}, we scaled the $\omega\gamma$s in Ref.~\cite{seuthe} by this factor.  

It is surprising that our strengths are several times higher than those from the previously reported direct $^{22}$Na($p,\gamma$) measurements~\cite{seuthe,stegmuller}.  However, those strengths were determined relative to resonance strengths from one experiment performed in 1990~\cite{seuthe}, also with implanted targets but with 2$\times$ the implantation energy and into a different substrate material.  We rastered the beam over the entire target, whereas the measurement of Ref.~\cite{seuthe} used a centered beam, which can degrade the target locally in ways that are difficult to characterize.  The gamma-ray energy window in that experiment was several MeV for most of the data, although two resonances were measured at peak yield with a high resolution detector.  Our energy windows were always narrow, with a maximum of a few tens of keV to incorporate only the relevant peak.  Along with a cosmic-ray anticoincidence system, our method employed full excitation functions integrated over proton energy and was largely independent of stopping-power estimations.  Our method required knowledge of only the total number of target atoms, determined from the target activity, and the requirement that the beam covered all target atoms, which we could monitor with the information from the raster.  The price paid for eliminating the dependence on the target distribution was the difficulty of determining the beam density.  This was accomplished experimentally using a $^{27}$Al coin target, and systematic effects were carefully considered.  In addition, our detector efficiency was determined with two radioactive sources and two resonances in the $^{27}$Al($p,\gamma$) reaction on two different substrates, spanning a gamma-ray energy scale of 1.3 to 11 MeV, in addition to calculation with detailed simulations.  Furthermore, the validity of our method has been confirmed with the $^{23}$Na($p,\gamma$) measurement.  A target was implanted, and result for the resonance strength are within 89 $\pm$ 24\% of the currently accepted value.  Although the error on this quantity is not negligible, it is not a factor of $\sim$ 3 we observe in the $^{22}$Na($p,\gamma$) resonance strengths.  Thus we are confident in the absolute values we have obtained for the strengths.

\section{Astrophysical Implications}\label{astro}

Using the resonance energies and total strengths shown in Table~\ref{table}, we calculated the contributions to the $^{22}$Na($p,\gamma)^{23}$Mg thermonuclear reaction rate under the narrow-resonance formalism.  Monte Carlo methods were employed, assuming symmetric (asymmetric) gaussian distributions for all measured resonances with symmetric (asymmetric) uncertainties.  For the proposed resonances at $E_p = 198$ and 232 keV, their distributions were taken to be the curves shown in Fig.~\ref{pdfs}.  $R_{\rm central}$ is defined as the 50\% quantile of the distribution of rates at a given temperature, and $R_{\rm upper}$ and $R_{\rm lower}$ are the 16\% and 84\% quantiles, respectively.

The proposed resonance at $E_p$ = 209 keV has not been included in the reaction rate, as its previously estimated strength~\cite{jenkins} is so weak that its contribution should be negligible.  In addition, the upper limit this work sets is conservative because of potential contributions from nearby resonances to the excitation function.  The strengths of the resonances at $E_p$ = 45 and 70 keV are too low to be measured directly due to the Coulomb barrier but have been included in the calculation of the reaction rate derived from Ref.~\cite{iliadispreprint3}, based on the ($^{3}$He,$d$) spectroscopic factors of Ref~\cite{schmidt}.

\begin{figure}
\hfil\scalebox{.45}{\includegraphics{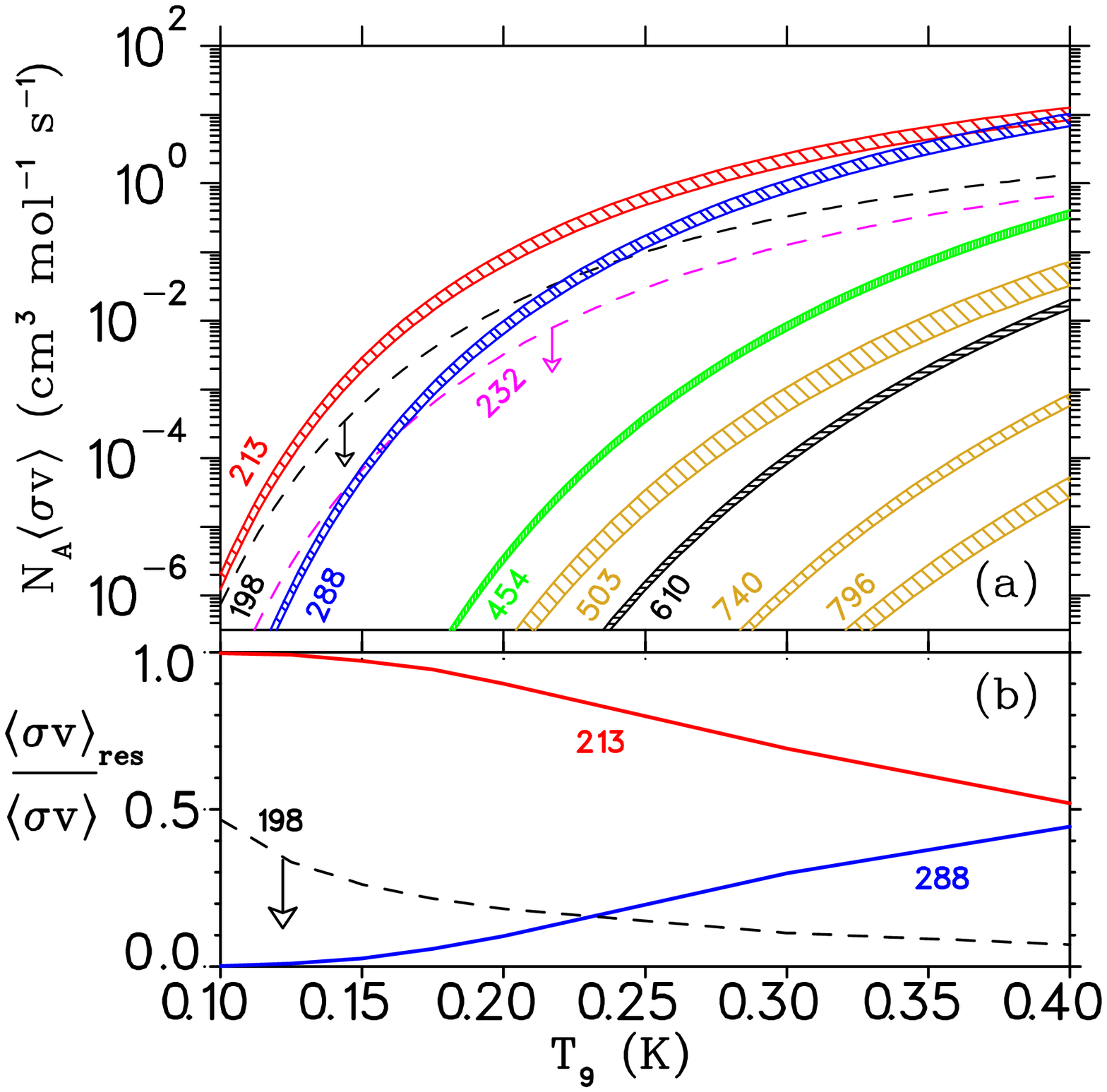}}\hfil
\caption{(color online) Thermonuclear $^{22}$Na($p,\gamma)^{23}$Mg reaction rate as a function of temperature.  Panel (a) shows contributions to the reaction rate from individual resonances labeled by $E_p$ (in keV), based on present measurements.  Hatched regions represent 68\% confidence levels (although the uncertainties are so small they may appear as solid lines), and dashed lines with arrows are 68\% confidence level upper limits.  Panel (b) shows the fractional contributions of selected resonances to the total rate, as calculated using the resonances illustrated in panel (a). }
\label{rxn}
\end{figure}%

In Fig.~\ref{rxn} (a), we show the individual contributions to the thermonuclear rate for $^{22}$Na($p,\gamma$), and panel (b) shows the relative contributions of selected resonances.  As one can see, any contribution from the resonance at $E_p$ = 198 keV is less than that from the resonance at $E_p$ = 213 keV for all temperatures of interest to novae.  Therefore, the resonance at $E_p = 198$ keV does not dominate the reaction rate in this region, as Jenkins {\it et al.}~\cite{jenkins} proposed it might.  Rather, the resonance at $E_p$ = 213 keV makes the most important contribution.  Also, at the higher temperatures around 0.4 GK, the contribution of the 288-keV resonance becomes significant.  The other resonance contributions are effectively negligible at nova temperatures.

\begin{figure}
\hfil\scalebox{.35}{\includegraphics{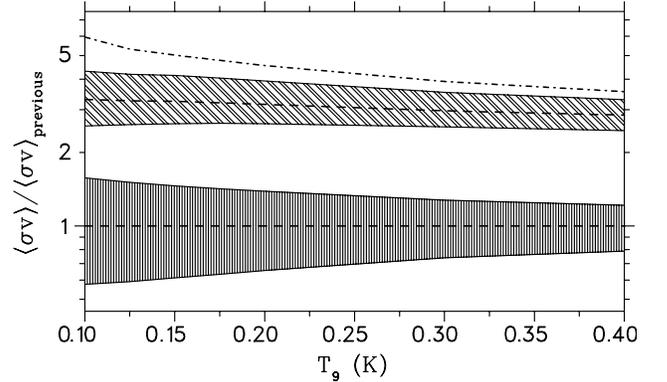}}\hfil
\caption{Ratio of present (slanted hatches) and previous (vertical hatches)~\cite{seuthe,stegmuller} to previous thermonuclear $^{22}$Na($p,\gamma)^{23}$Mg reaction rate as a function of temperature.  Hatched areas represent 68\% $C.L.$ error bands and dashed lines represent the central value, each relative to the previous central value.  Both previous and present resonance strengths are listed in Table~\ref{table}.  Including the distributions for the resonances at $E_p = 198$ and 232 keV increases the present upper limit to the dot-dashed line.  The upper limit for the 198-keV resonance from Jenkins {\it et al.}~\cite{jenkins} has not been included in the previous rate.}
\label{past}
\end{figure}%

Fig.~\ref{past} illustrates the total reaction rate relative to previous direct measurements~\cite{seuthe,stegmuller}, showing that our rate is inconsistent with previous work at all temperatures.  The dashed line and hatched region in this figure do not include the distributions for proposed resonances at $E_p=$ 198 and 232 keV.  The 232-keV resonance makes an insignificant contribution at all temperatures of interest to novae.  However, although the 198-keV resonance was unobserved and an upper limit has been placed on its strength, including its potential contributions to the reaction rate has a non-negligible effect at low temperatures.  Including the proposed resonances increases the upper limit of the reaction rate to the dot-dashed line shown in Fig.~\ref{past}.  Because of this difference, we have calculated the reaction rate for each of these two separate cases, and the values are shown in Table~\ref{temp}.

\begin{table*}
\caption{Table of the thermonuclear $^{22}$Na($p,\gamma)^{23}$Mg reaction rate, $R$, as a function of temperature, $T$, determined using the energies and strengths given in Table~\ref{table}.  Two distributions of rates were calculated, one of which includes the probability density functions from resonances at $E_p = 198$ and 232 keV and one which does not.  $R_{\rm central}$ is the 50\% quantile of the distribution of rates at a given temperature, and $R_{\rm upper}$ and $R_{\rm lower}$ are the 16\% and 84\% quantiles, respectively.  The units of the rate are cm$^3$mol$^{-1}$s$^{-1}$. }
\begin{ruledtabular}
\begin{tabular}{c|ccc|ccc}
&\multicolumn{3}{c|}{R (not including ``198, 232"\footnotemark[1])}&\multicolumn{3}{c}{R (including ``198, 232"\footnotemark[1])}\\
$T_9$ (K) & $R_{\rm central}$&$R_{\rm lower}$&$R_{\rm upper}$&$R_{\rm central}$&$R_{\rm lower}$&$R_{\rm upper}$\\
\colrule
0.01	&	2.0$\times 10^{-30}$	&	2.7$\times 10^{-31}$	&	1.5$\times 10^{-29}$	&	2.0$\times 10^{-30}$	&	2.7$\times 10^{-31}$	&	1.5$\times 10^{-29}$	\\
0.02	&	5.5$\times 10^{-20}$	&	1.9$\times 10^{-20}$	&	1.6$\times 10^{-19}$	&	5.5$\times 10^{-20}$	&	1.9$\times 10^{-20}$	&	1.6$\times 10^{-19}$	\\
0.03	&	2.6$\times 10^{-16}$	&	1.3$\times 10^{-16}$	&	5.6$\times 10^{-16}$	&	2.6$\times 10^{-16}$	&	1.3$\times 10^{-16}$	&	5.6$\times 10^{-16}$	\\
0.04	&	4.6$\times 10^{-14}$	&	2.0$\times 10^{-14}$	&	1.1$\times 10^{-13}$	&	4.6$\times 10^{-14}$	&	2.1$\times 10^{-14}$	&	1.1$\times 10^{-13}$	\\
0.05	&	1.4$\times 10^{-12}$	&	6.4$\times 10^{-13}$	&	3.1$\times 10^{-12}$	&	1.4$\times 10^{-12}$	&	6.4$\times 10^{-13}$	&	3.1$\times 10^{-12}$	\\
0.06	&	1.4$\times 10^{-11}$	&	7.1$\times 10^{-12}$	&	2.8$\times 10^{-11}$	&	1.5$\times 10^{-11}$	&	7.7$\times 10^{-12}$	&	2.9$\times 10^{-11}$	\\
0.07	&	1.8$\times 10^{-10}$	&	1.3$\times 10^{-10}$	&	2.5$\times 10^{-10}$	&	2.7$\times 10^{-10}$	&	2.0$\times 10^{-10}$	&	3.6$\times 10^{-10}$	\\
0.08	&	6.2$\times 10^{-09}$	&	4.7$\times 10^{-09}$	&	8.2$\times 10^{-09}$	&	9.7$\times 10^{-09}$	&	7.2$\times 10^{-09}$	&	1.3$\times 10^{-08}$	\\
0.09	&	1.3$\times 10^{-07}$	&	1.0$\times 10^{-07}$	&	1.8$\times 10^{-07}$	&	2.0$\times 10^{-07}$	&	1.5$\times 10^{-07}$	&	2.6$\times 10^{-07}$	\\
0.10	&	1.6$\times 10^{-06}$	&	1.2$\times 10^{-06}$	&	2.1$\times 10^{-06}$	&	2.2$\times 10^{-06}$	&	1.7$\times 10^{-06}$	&	2.8$\times 10^{-06}$	\\
0.15	&	2.3$\times 10^{-03}$	&	1.9$\times 10^{-03}$	&	3.0$\times 10^{-03}$	&	2.9$\times 10^{-03}$	&	2.4$\times 10^{-03}$	&	3.6$\times 10^{-03}$	\\
0.2	&	8.5$\times 10^{-02}$	&	7.1$\times 10^{-02}$	&	1.1$\times 10^{-01}$	&	1.0$\times 10^{-01}$	&	8.4$\times 10^{-02}$	&	1.2$\times 10^{-01}$	\\
0.3	&	3.1$\times 10^{+00}$	&	2.7$\times 10^{+00}$	&	3.7$\times 10^{+00}$	&	3.5$\times 10^{+00}$	&	3.0$\times 10^{+00}$	&	4.1$\times 10^{+00}$	\\
0.4	&	1.9$\times 10^{+01}$	&	1.7$\times 10^{+01}$	&	2.2$\times 10^{+01}$	&	2.1$\times 10^{+01}$	&	1.8$\times 10^{+01}$	&	2.4$\times 10^{+01}$	\\
0.5	&	5.9$\times 10^{+01}$	&	5.1$\times 10^{+01}$	&	6.7$\times 10^{+01}$	&	6.3$\times 10^{+01}$	&	5.4$\times 10^{+01}$	&	7.1$\times 10^{+01}$	\\
0.6	&	1.3$\times 10^{+02}$	&	1.1$\times 10^{+02}$	&	1.4$\times 10^{+02}$	&	1.3$\times 10^{+02}$	&	1.2$\times 10^{+02}$	&	1.5$\times 10^{+02}$	\\
0.7	&	2.3$\times 10^{+02}$	&	2.0$\times 10^{+02}$	&	2.5$\times 10^{+02}$	&	2.4$\times 10^{+02}$	&	2.1$\times 10^{+02}$	&	2.6$\times 10^{+02}$	\\
0.8	&	3.6$\times 10^{+02}$	&	3.2$\times 10^{+02}$	&	4.0$\times 10^{+02}$	&	3.7$\times 10^{+02}$	&	3.3$\times 10^{+02}$	&	4.1$\times 10^{+02}$	\\
0.9	&	5.4$\times 10^{+02}$	&	4.8$\times 10^{+02}$	&	5.9$\times 10^{+02}$	&	5.5$\times 10^{+02}$	&	5.0$\times 10^{+02}$	&	6.0$\times 10^{+02}$	\\
1.0	&	7.5$\times 10^{+02}$	&	6.8$\times 10^{+02}$	&	8.1$\times 10^{+02}$	&	7.6$\times 10^{+02}$	&	6.9$\times 10^{+02}$	&	8.3$\times 10^{+02}$	\\
1.5	&	2.2$\times 10^{+03}$	&	2.1$\times 10^{+03}$	&	2.4$\times 10^{+03}$	&	2.2$\times 10^{+03}$	&	2.1$\times 10^{+03}$	&	2.4$\times 10^{+03}$	\\
2.0	&	3.9$\times 10^{+03}$	&	3.6$\times 10^{+03}$	&	4.2$\times 10^{+03}$	&	3.9$\times 10^{+03}$	&	3.6$\times 10^{+03}$	&	4.2$\times 10^{+03}$	\\

\end{tabular}
\end{ruledtabular}
\footnotetext[1]{``198, 232" denotes resonances at $E_p = 198$ and 232 keV.}
\label{temp}
\end{table*}

\begin{table}
\caption{$^{22}$Na yields from novae on ONe white dwarfs of various masses computed with SHIVA~\cite{jose98}.  The total mass ejected is given for each case, along with  the mass fractions of $^{22}$Na obtained using the central value of the previous~\cite{seuthe,stegmuller} and both present $^{22}$Na($p,\gamma)^{23}$Mg reaction rates compiled in Table~\ref{temp}.  The factor given is the ratio of previous to respective present amounts of $^{22}$Na ejected.}
\begin{ruledtabular}
\begin{tabular}{lccc}
 & $1.15M_{\odot}$&$1.25M_{\odot}$&$1.35M_{\odot}$\footnotemark[1]\\
\colrule
 M$_{\rm eject}$ (g)    &  4.9$\times 10^{28}$   &  3.8$\times 10^{28}$    & 9.0$\times 10^{27}$    \\
\colrule
Previous\footnotemark[2]&1.6$\times 10^{-4}$&1.9$\times 10^{-4}$&5.9$\times 10^{-4}$ \\
\hline
Present\footnotemark[3]&8.8$\times 10^{-5}$&1.1$\times 10^{-4}$&4.1$\times 10^{-4}$ \\
Factor&1.8&1.8&1.4 \\
\\
Present\footnotemark[4] &  7.8$\times 10^{-5}$ &               1.0$\times 10^{-4}$     &           3.8$\times 10^{-4}$\\
Factor & 2.0          &           1.9                       & 1.5\\
\end{tabular}
\end{ruledtabular}
\footnotetext[1]{Statistically, there should be more novae of $1.15M_{\odot}$ and $1.25M_{\odot}$ than those hosting $1.35M_{\odot}$, due to the stellar mass function of the progenitors. }
\footnotetext[2]{Rate derived using energies and strengths from Ref.~\cite{seuthe,stegmuller}.}
\footnotetext[3]{Not including ``198, 232" from Table~\ref{temp}.}
\footnotetext[4]{Including ``198, 232" from Table~\ref{temp}.}

\label{table_jose}
\end{table}

We can anticipate the general ramifications of the new rate on expected nucleosynthesis of $^{22}$Na in ONe novae using post-processing network calculations because the total energy generation is not affected appreciably by the $^{22}$Na($p,\gamma)$ reaction.  Based on the one-zone calculations of Ref.~\cite{hix}, a specific model indicates that the production of $^{22}$Na is related inversely to the $^{22}$Na($p,\gamma)$ reaction rate.  The authors of Ref.~\cite{iliadis} also varied the rate by a factor of two using various one-zone models of novae to extract the effect.  Using the information given in these references and our change in the reaction rate, the estimated abundance of $^{22}$Na in novae is expected to be reduced by factors of 2 to 3 of what was previously expected, depending on white dwarf mass and composition.  This will directly affect the expected flux of $^{22}$Na gamma rays observed using orbiting gamma-ray telescopes. 

In addition, the impact of our $^{22}$Na$(p,\gamma)$ reaction rate on the amount of $^{22}$Na ejected during nova outbursts has been tested through a series of hydrodynamic simulations:  three evolutionary sequences of nova outbursts hosting ONe white dwarfs of 1.15, 1.25 and $1.35 M_\odot$ have been computed with the spherically symmetric, Lagrangian, hydrodynamic code SHIVA, extensively used in the modeling of such explosions (see Ref.~\cite{jose98}, for details).  Results have been compared with those obtained in three additional hydrodynamic simulations, for the same white dwarf masses described above and same input physics except for the $^{22}$Na$(p,\gamma)$ rate, which was derived from Refs.~\cite{stegmuller,seuthe}.  The network used for additional reaction rates is the relevant subset of that used in Ref.~\cite{jordiref}.  The estimated $^{22}$Na yields (mass-averaged mass fractions in the overall ejected shells) are listed in Table~\ref{table_jose}, which clearly shows that the impact of the central value of new rate roughly translates into lower $^{22}$Na abundances by a factor up to $\sim$ 2 with respect to previous estimates.  This, in turn, directly affects the chances to potentially detect the 1275-keV gamma-ray line from $^{22}$Na decay, decreasing the maximum detectability distances by a factor $\sim 1.4$.  The inclusion of the 198- and 232-keV distributions in the rate does not appreciably alter this factor.  

The results from one-zone post-processing network calculations and full hydrodynamic simulations using SHIVA are complementary.  The post-processing approach mimics the processes that occur in the deepest envelope layers, whereas the hydrodynamic simulations average the yields over all ejected shells.  Convection also plays a critical role, supplying freshly, unburned material from external shells into the innermost one (and vice versa), and these effects cannot be simulated in a post-processing framework.  As a result of the more realistic physics in the hydrodynamic model, the composition of the innermost shell is diluted by the compositions of the outermost ones.  On the other hand, the post-processing calculations cover various nova models, a wider range of nova masses and compositions, and show that the correlation between the $^{22}$Na($p,\gamma)$ rate and $^{22}$Na production is robust even when other reaction rates are simultaneously varied.  It seems reasonable to assume that the magnitude of the dilution from the hydrodynamic models ($\approx 25 \%$) applies generally to all of the post-processing results.

\section{Conclusions}

We have measured the resonance strengths, energies, and branches of the $^{22}$Na($p,\gamma)^{23}$Mg reaction directly and absolutely.  Our method improved upon past measurements in several ways.  The use of integrated yields makes the results independent of absolute stopping power calculations and is far more robust than using peak yields.  We also utilized isotopically-pure, implanted targets that demonstrated nearly zero loss during bombardment.  HPGe detectors exhibit excellent energy resolution, providing the ability to use narrow energy windows, and anticoincidence shields enabled suppression of the cosmic ray background.  Absolute detector efficiency was also vital, which we determined by fusing measurement and simulation.  Finally, the rastering of the beam across the target not only aided in maintaining target integrity, but also removed the requirement of detailed knowledge of the target distribution.  A determination of the beam density was mandatory and was ascertained by both measurement and modeling.  As a consequence of the aforementioned points, our results should be substantially more reliable than previous measurements.

By exploiting these advantages, our measurement has shown that four previously measured resonance strengths are 2.4 to 3.2 times higher than previously reported~\cite{seuthe,stegmuller}.  Jenkins {\it et al.} also proposed that a new $^{22}$Na($p,\gamma$) resonance with $E_p =$ 198 keV could dominate this reaction rate and have consequences for novae~\cite{jenkins}.  We have demonstrated that this is not the case, and that the main contributions arise from the resonance at $E_p$ = 213 keV.  As a result of the higher resonance strengths, the estimated flux of $^{22}$Na gamma rays from novae is expected to be about a factor of 2 less than what was previously expected, determined by using both post-processing network calculations and hydrodynamic simulations.  The lack of observational evidence of $^{22}$Na in the cosmos is consistent with the previous reaction rate; however, the present rate makes detection $\sim$ 1.4 times more difficult to detect.  

\acknowledgments

We gratefully acknowledge the contributions of G. Harper, D. I. Will, D. Short, B. M. Freeman, K. Deryckx, and the technical staffs at CENPA and TRIUMF, as well as the Athena Cluster. This work was supported by the U.S. Department of Energy under contract No. DE-FG02-97ER41020, the Natural Science and Engineering Research Council of Canada, and the National Research Council of Canada. 


\end{document}